\documentstyle[emulateapj]{article}

\lefthead{Alonso-Herrero et al.}
\righthead{The interacting galaxy Arp~299}

\begin{document}
\title{Extreme Star Formation in the Interacting Galaxy Arp~299 
(IC~694 + NGC~3690)
\footnote{Based
on observations with the NASA/ESA Hubble
Space Telescope, obtained at the Space Telescope Science Institute, which
is operated by the Association of Universities for Research in Astronomy,
Inc. under NASA contract No. NAS5-26555. Some of the observations
reported in this paper were obtained with the
Multiple Mirror Telescope, which is operated jointly by the Smithsonian
Astrophysical Observatory and the University of Arizona.}}

\author{Almudena Alonso-Herrero, George H. Rieke, Marcia J. Rieke}
\affil{Steward Observatory, The University of Arizona, Tucson, AZ 85721}

\and

\author{Nick Z. Scoville}
\affil{California Institute of Technology, Pasadena, CA 91125}

\begin{abstract}
We present a comprehensive study of the star-formation properties of 
the infrared luminous galaxy Arp~299 (IC~694 + NGC~3690). The 
observations include {\it HST}/NICMOS imaging and MMT optical and 
near-infrared
spectroscopy together with {\it HST} archival data. We correct the 
galaxy parameters for extinction and use the results as boundary 
conditions for evolutionary starburst models. 

These models and other arguments show that Arp 299 has been going through 
a broad variety of interaction-induced star formation for the last 
$\sim$ 15 Myr. In addition to nuclear starbursts in the two 
colliding galaxies, 
there are nearby luminous star forming regions that may be the result 
of molecular clouds breaking up and starting to form massive stars as 
they approach the nuclear potential. Two regions near the interaction 
region have very recently formed massive stars ($\sim$ 4 Myr old). 
One of these regions in particular appears to have undergone a starburst 
of very short duration, in contrast with the 5-10 Myr durations that 
are typical of nuclear events. These regions will probably form 
gravitationally bound new dwarf galaxies, although they are close 
enough to the more massive original galaxies that they will probably 
eventually be subsumed into them. In addition, we find 19 H\,{\sc ii} 
regions at least as energetic as 30 Doradus, and 21 older star 
clusters that likely are the products of similar H\,{\sc ii} regions formed 
in the past. This population of supergiant H\,{\sc ii} regions is 
unprecedented 
in normal galaxies and emphasizes that the effects of the interaction 
propagate through the entire pair of galaxies. 

\end{abstract}


\section{INTRODUCTION}
Ultraluminous infrared galaxies that rival quasars in energy output were
first identified nearly thirty years ago (Rieke \& Low 1972), and were
shown by {\it IRAS} fifteen years ago to exist in numbers comparable to those
of quasars in the nearby Universe (e.g., Sanders et al. 1988). The 
exact nature of these extreme objects has been of great interest and 
debate since. Among the ultraluminous infrared class 
($L_{\rm IR} > 10^{12}\,$L$_\odot$) it is found that a large 
percentage are interacting/merging systems containing active galactic 
nuclei. The process of merging with accompanying super-starbursts 
appears to be an important stage in galaxy evolution, possibly even 
converting spiral galaxies into ellipticals. Sanders et al. (1988) 
suggested that the infrared luminous phase is the initial stage 
for the appearance of a quasar (see Sanders \& Mirabel 1996 for a 
recent review). 

\begin{deluxetable}{lccc}
\tablefontsize{\footnotesize}
\tablewidth{12cm}
\tablecaption{Log of the NICMOS observations}
\tablehead{\colhead{Camera} & \colhead{Filter} &
\colhead{$t_{\rm exp}$}& \colhead{Field of view}}
\startdata
NIC3   & F164N & 1024 & $51.2\arcsec \times 51.2\arcsec$ (IC~694+NGC~3690)\\
NIC3   & F166N & 1024 & $51.2\arcsec \times 51.2\arcsec$ (IC~694+NGC~3690)\\
NIC2   & F160W & 208 & $19.5\arcsec \times 19.5\arcsec$ (IC~694), 
$19.5\arcsec \times 19.5\arcsec$ (NGC~3690) \\
NIC2   & F222M & 528 & $19.5\arcsec \times 19.5\arcsec$ (IC~694), 
$19.5\arcsec \times 19.5\arcsec$ (NGC~3690)\\
NIC2   & F237M & 864 & $19.5\arcsec \times 19.5\arcsec$ (IC~694), 
$19.5\arcsec \times 19.5\arcsec$ (NGC~3690) \\
NIC2   & F187N & 320  & $19.5\arcsec \times 19.5\arcsec$ (IC~694), 
$19.5\arcsec \times 19.5\arcsec$ (NGC~3690)\\
NIC2   & F190N & 320 & $19.5\arcsec \times 19.5\arcsec$ (IC~694), 
$19.5\arcsec \times 19.5\arcsec$ (NGC~3690)\\
NIC2   & F212N & 2560 & $19.5\arcsec \times 19.5\arcsec$ (IC~694), 
$19.5\arcsec \times 19.5\arcsec$ (NGC~3690)\\
NIC2   & F215N & 2560 & $19.5\arcsec \times 19.5\arcsec$ (IC~694), 
$19.5\arcsec \times 19.5\arcsec$ (NGC~3690)\\
NIC1   & F110M & 256 & $42.6 \arcsec \times 21.0\arcsec$ (IC~694 + NGC~3690)
8 image mosaic\\
\enddata
\end{deluxetable}

To understand these objects requires observations at high physical 
resolution. However, because their space density is low at the current 
epoch, there are few nearby examples where such resolution is possible. 
Arp~299 (NGC~3690/IC~694 or Mrk~171) is of high infrared luminosity 
($L_{\rm IR} = 5 \times 10^{11}\,$L$_\odot$), placing it among the 
luminous infrared galaxies, but also is one of the nearest 
examples of interacting starburst galaxies (distance $D = 42\,$Mpc 
for $H_0 = 75\,$km s$^{-1}$ Mpc$^{-1}$).

Within Arp~299, a number of bright infrared and radio sources appear 
to be the regions of star formation and/or nuclear activity 
responsible for the large overall luminosity (see Gehrz, Sramek, 
\& Weedman 1983; 
Wynn-Williams et al. 1991 and references therein){\footnote {We 
use the notation introduced by Gehrz et al. (1983) for the interacting
pair of galaxies Arp~299: the nucleus of IC~694 (eastern component) is
source A, and the sources in NGC~3690 (western component)  are B1,
B2, C and C$^\prime$ (see also Wynn-Williams et al. 1991 and Alonso-Herrero,
Rieke, \& Rieke 1998).}}. As in other luminous and ultraluminous galaxies, 
there is a high 
concentration of molecular hydrogen within relatively small
regions around these sources. From CO maps, Sargent \& Scoville (1991) 
estimated densities of molecular gas 
$\simeq 8 \times 10^5\,{\rm M}_\odot$ pc$^{-2}$ in
IC~694, $\simeq 3 \times 10^4\,{\rm M}_\odot$ pc$^{-2}$ in components
B1 and B2 of NGC~3690 and $\simeq 2 \times 10^4\,{\rm M}_\odot$ pc$^{-2}$
at the interface of both galaxies (region C+C'). These general 
characteristics are consistent with numerical simulations of 
collisions between gas-rich galaxies
(see Barnes \& Hernquist 1996 and references therein) that show 
how interactions lead to transportation of large quantities of
molecular gas into the centers of galaxies. This concentration of gas 
leads to a strong burst of star-formation and may activate an AGN.

However, these properties make the centers of activity in 
luminous and ultraluminous 
infrared galaxies virtually impenetrable in the visible and 
ultraviolet. We report near infrared {\it HST}/NICMOS images that 
penetrate much of the interstellar extinction and are at an 
unprecedented level of resolution for this spectral region, up to 
0.1"\, corresponding to 20 pc. We also use WFPC2 images of the 
system together with ground-based optical and infrared spectroscopy to 
derive the star formation history. These new data are placed
in the context of the characteristics of the galaxy measured in
other spectral regions, as summarized briefly in the preceding paragraph.

\section{OBSERVATIONS}
\subsection{NICMOS observations}
{\it HST}/NICMOS observations of Arp~299  were
obtained on November 4 1997 using all three cameras. In Table~1 we
give the log of observations, where column~(2) is the filter, column~(3)
is the integration time and column~(4) is the field of view of the
images. The observations were obtained in  
a spiral dither with a 5.5 pixel spacing and with two, three or
four  positions. The plate scales for NIC1, NIC2 and NIC3 are
0.045"\,pixel$^{-1}$, 0.076"\,pixel$^{-1}$ and 0.20"\,pixel$^{-1}$
respectively.

The images were reduced with routines from  the package
{\it NicRed} (McLeod 1997).  The main steps in the data reduction involve 
subtraction of the first readout, dark current subtraction on a 
readout-by-readout basis, correction for linearity and cosmic
ray rejection (using fullfit), and flatfielding. 
Darks with sample sequences and exposure times
corresponding to those of our observations were obtained from other 
programs close in time to ours. Usually between 10 and 20 darks were 
averaged together (after the subtraction of the first readout) for a 
given sample sequence. However, in the case of F110M we got better 
removal of the artifacts using a dark+sky image generated by median 
combining all the dithered galaxy images for this filter.
Flatfield images were constructed from  on-orbit data. 
Since our images were obtained after
August 1997, no correction for the pedestal effect was necessary.
For more specific details on the reduction of the NIC3 images see
Alonso-Herrero et al. (1998).

After removal of artifacts, the dithered galaxy images 
were registered to a common position using 
fractional pixel offsets and cubic spline interpolation, and combined to 
produce the final images. The NIC1 F110M images were mosaiced. 
Since there are no bright common sources in the overlapping
regions of the images, we used the world coordinate information contained
in the header to compute the offsets. All the NICMOS images 
were rotated to the usual orientation north up, east to the left.

Prior to flux-calibrating the images, we measured the background on 
blank corners of the images. Although this interacting pair of galaxies is
known to be extended beyond the field of the NIC2 images, the background
values measured in the NIC2 F222M filter image
are in excellent agreement with the background measurements taken during
the Servicing Mission Observatory Verification (SMOV) program.
The flux calibration was performed using the conversion factors based on
measurements of the standard star P330-E (Rieke 1999). The fully-reduced 
narrow-band images (NIC2 F187N, F190N, F212N and F215N, and NIC3 F164N 
and F166N) were first background subtracted, then flux calibrated, and 
finally shifted to a common position. The adjacent continuum image was 
subtracted from the corresponding line + continuum image to produce 
the final continuum-free line images.

\begin{figure*}
\figurenum{1}
\plotfiddle{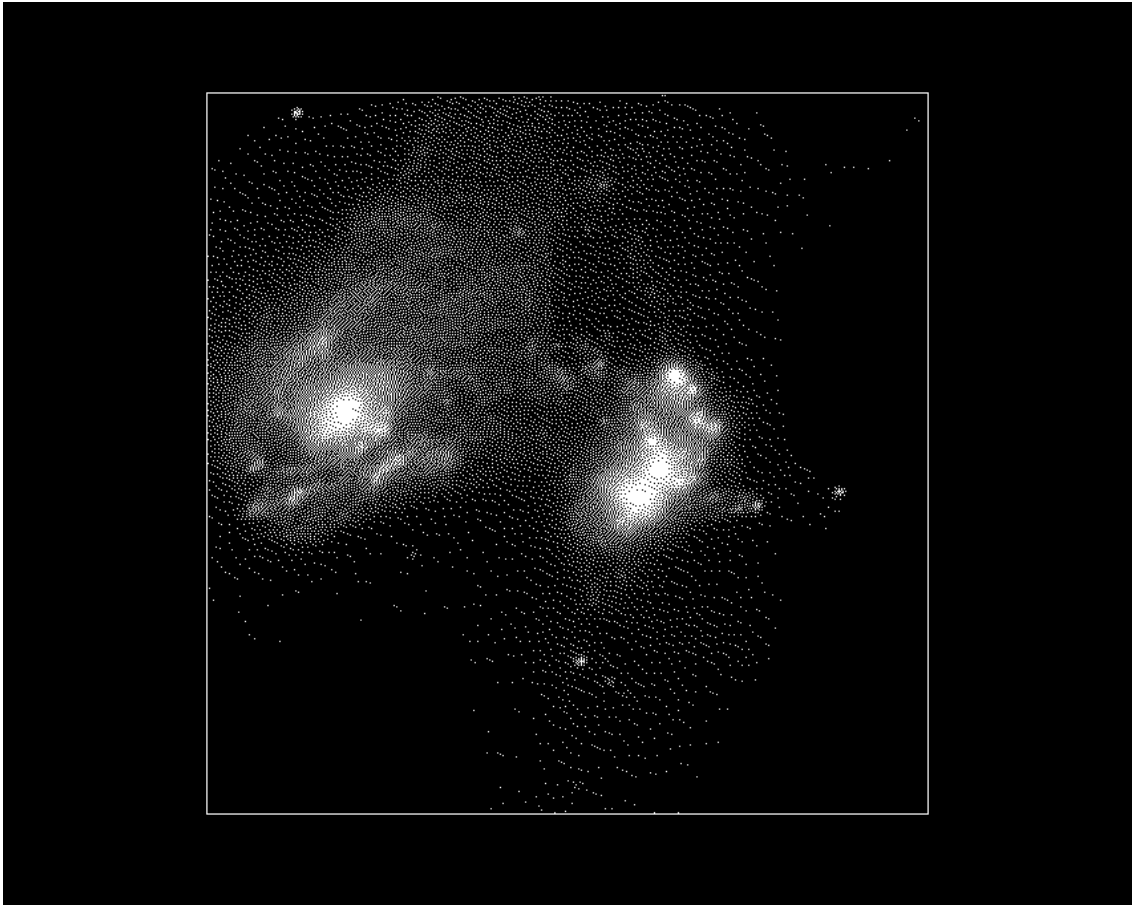}{425pt}{0}{50}{50}{-130}{0}
\caption{NIC3 F166N image ([Fe\,{\sc ii}]$1.644\,\mu$m + continuum)
of the interacting system
Arp~299 on a logarithmic scale. The total field of view is $51.2\arcsec
\times 51.2\arcsec$. Orientation is north up, east to the left.}
\end{figure*}

Figure~1 shows the NIC3 F166N 
image of the pair of interacting
galaxies. The NIC3 F166N filter at the redshift of Arp~299 contains 
the emission line [Fe\,{\sc ii}]1.644\,$\mu$m. The total field of view 
is $51.2\arcsec \times 51.2\arcsec$. The corresponding images of the 
continuum subtracted [Fe\,{\sc ii}]$1.644\,\mu$m line emission of the 
system can be found in Alonso-Herrero et al. (1998). Figure~2a and 
Figure~2b show both components through different filters as indicated 
in the figure caption. The field of view of all these images (except 
for the NIC1 F110M inset in Figure~2a) is $19.5\arcsec \times 19.5\arcsec$. 
The NIC2 F222M
and F237M images are similar in morphology and are not shown here.

\begin{figure*}
\figurenum{2a}
\plotfiddle{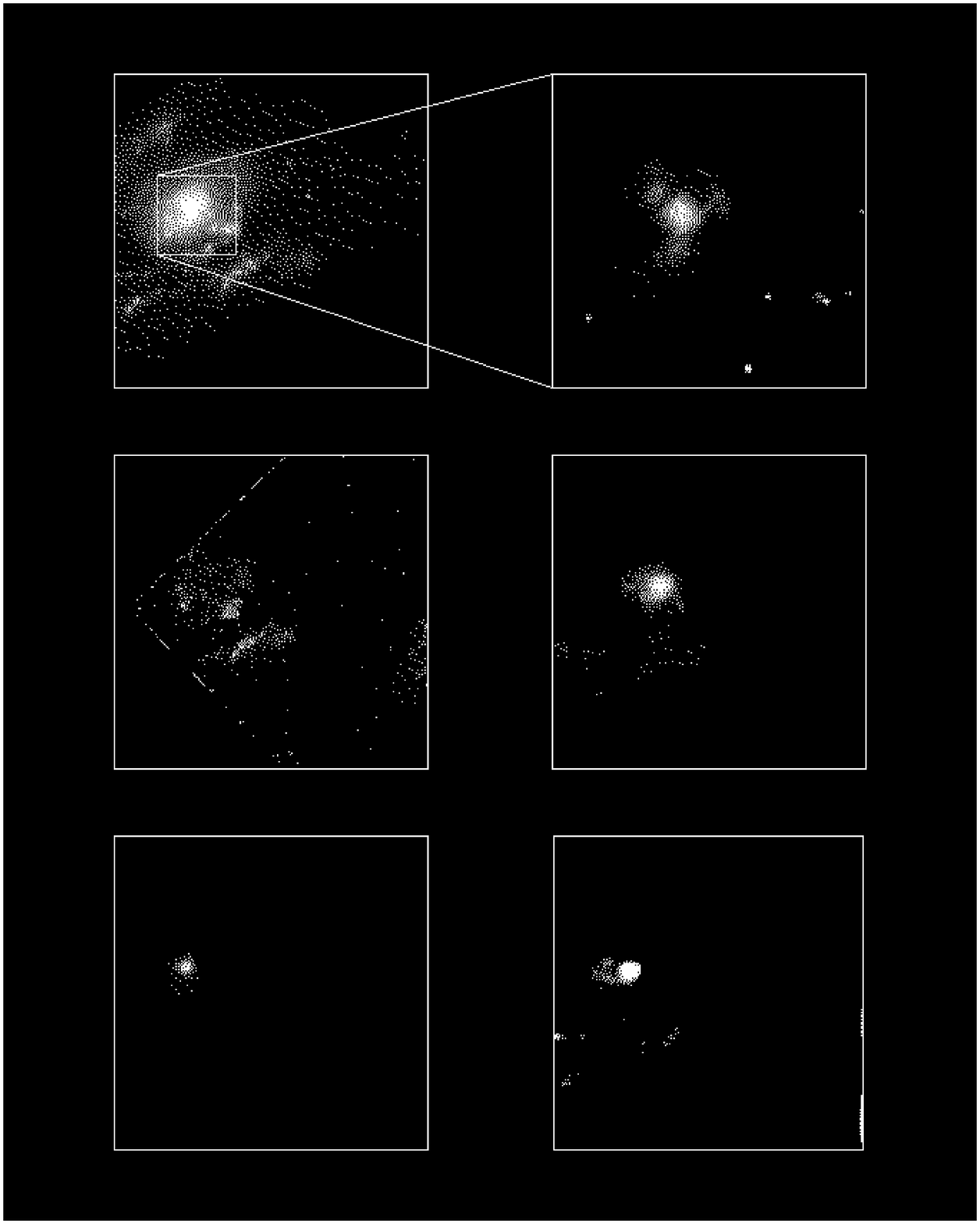}{425pt}{0}{50}{50}{-130}{50}
\caption{IC~694. Top panels, left: 
NIC2 F160W ($H$-band continuum),  and right: 
NIC1 F110M (continuum at $1.1\,\mu$m) right. Middle panels, 
left: optical WFPC2 F606W,  and right: 
continuum-subtracted NIC3 F166N ([Fe\,{\sc ii}]$1.644\,\mu$m).
Bottom panels, left: continuum
subtracted NIC2 F215N (H$_2$ at $2.12\,\mu$m), and right:
continuum-subtracted NIC2 F190N (Pa$\alpha$). Orientation
is north up, east to the left. The field of view of all the images
except the inset is $19.5\arcsec \times 19.5\arcsec$.}
\end{figure*}

\begin{figure*}
\figurenum{2b}
\plotfiddle{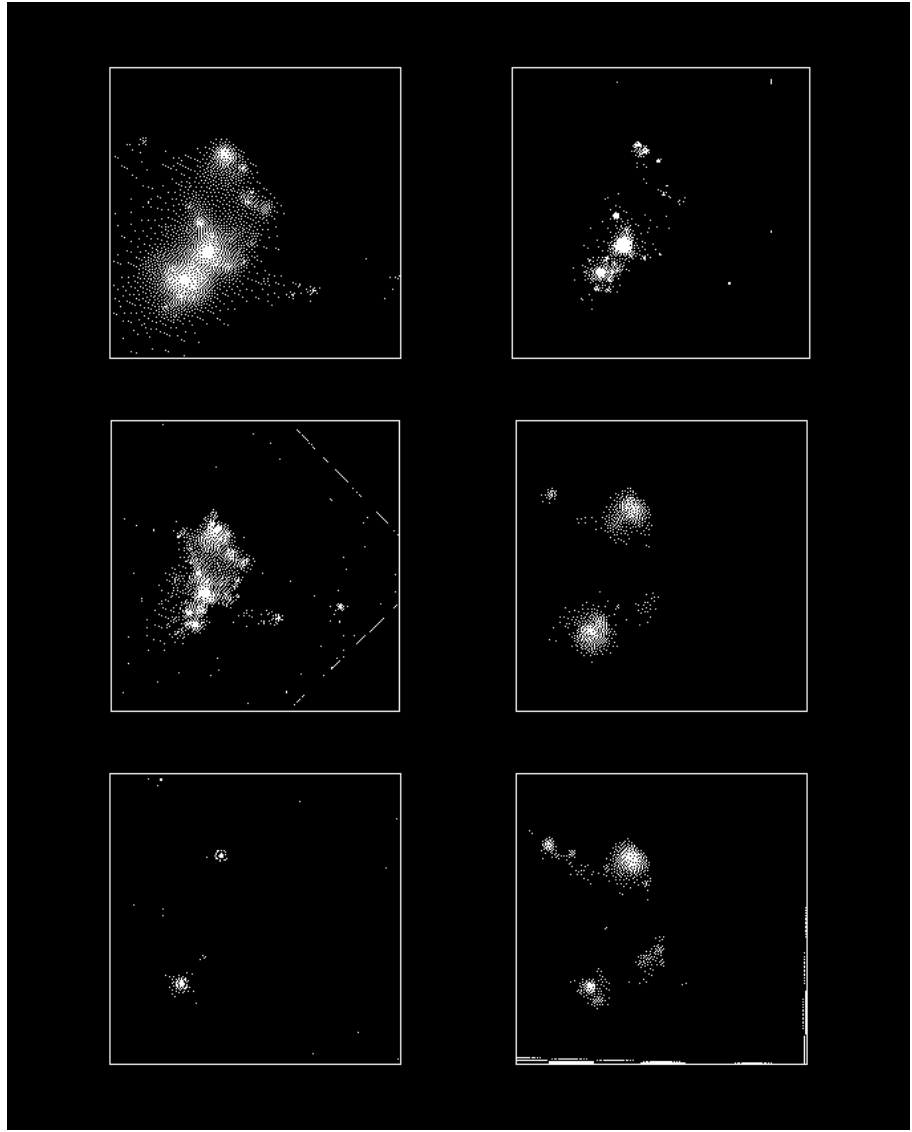}{425pt}{0}{50}{50}{-130}{50}
\caption{Same as Figure~2a but for NGC~3690.}
\end{figure*}

\begin{deluxetable}{lcccccc}
\tablefontsize{\footnotesize}
\tablewidth{13cm}
\tablecaption{Low-resolution $K$-band infrared spectroscopy
for a $1.2\arcsec \times 1.2\arcsec$ extraction aperture.}
\tablehead{\colhead{Source}&
\colhead{$f({\rm H}_2$)} & \colhead{EW(H$_2$)} &
\colhead{$f({\rm Br}\gamma$)}&
\colhead{EW(Br$\gamma$)} & \colhead{(CO)$_{\rm sp}$} &
\colhead{(CO)$_{\rm ph}$}\\
\colhead{}& \colhead{(erg cm$^{-2}$ s$^{-1}$)} & \colhead{(\AA)} &
\colhead{(erg cm$^{-2}$ s$^{-1}$)} & \colhead{(\AA)} & & }
\startdata
B1 & $9.52 \times 10^{-15}$ & 10.2 & $8.56\pm 0.5 \times 10^{-15}$
& $9.0\pm 0.5$ & 0.14 & 0.07  \\
B2 & $<2.4\times 10^{-15}$ &  $<4$ & $<3.1 \times 10^{-15}$ & $<5$
& 0.31 &0.17 \\
C  & $2.95 \times 10^{-15}$ & 5.1 & $2.43 \pm 0.1 \times 10^{-14}$ &
$41\pm 2$ & \nodata & \nodata \\
\enddata
\end{deluxetable}

\subsection{Optical and ultraviolet images from the {\it HST} archive}
A Wide Field Planetary Camera 2 (WFPC2) image of Arp~299 was retrieved 
from the {\it HST} data archive. This image was taken through the F606W 
filter on September 17 1994 with a total integration time of 500\,s and 
spatial
sampling of 0.046\arcsec \, pixel$^{-1}$. Standard pipeline reduction 
procedures were applied. The image was rotated to the usual 
orientation north up, east to the left. Since a single integration 
was taken for this galaxy, the cosmic ray
removal is problematic. The {\it cosmicray} task in {\sc iraf} successfully 
eliminated the most point-like hits but artifacts of extended hits 
remain in the data. The image is presented in Figure~2a and 2b, with 
the field of view matching that of NIC2. 

We also retrieved UV FOC images from the {\it HST} archive for each of
the components in the system, taken through the
UV F220W filter with integration time of 897\,s for each component of 
the pair.
Standard pipeline reduction procedures were used. 
Again the images were rotated to match the orientation of the NICMOS
images. The plate scale of these images is 0.022\arcsec\,pixel$^{-1}$.
The image of NGC~3690 was part of the study by Meurer et al. (1995) 
of the UV properties of starburst galaxies.

\subsection{Ground-based H$\alpha$ imaging}
CCD H$\alpha$ narrow-band images (on-line and off-line) of Arp~299 were
obtained from the La Palma Observatory
archive. The images were taken with the 1.5\,m JKT telescope 
on January 21 1990 with a CCD-RCA2 detector. The integration time for
the continuum+line and continuum images was 1,800\,s each. The reduction 
steps involved bias removal, dark current subtraction, and
flat-fielding with both dome and sky flats. The plate scale
of the images was 0.4\arcsec\,pixel$^{-1}$, providing a total field of
view of $130\arcsec \times 210\arcsec$. 

Since no images of suitable standard stars 
are provided by the La Palma archive, the final continuum-subtracted
H$\alpha$ image was calibrated with the spectroscopic data from
Armus et al. (1989) and Gehrz et al. (1983). The latter reference
provides fluxes through an effective aperture
diameter of 5 \arcsec. The former authors provide
the H$\alpha$ fluxes for components
A and C of the interacting pair through a $2\arcsec \times 4.5\arcsec$
aperture. Because of difficulties in setting the slit accurately on the 
heavily obscured sources, there are significant uncertainties in the line 
measurements. For example, Armus et al. (1989) report source C to have 
twice as much flux
as source A, whereas our H$\alpha$ continuum-subtracted
image indicates that source C is 
four times as bright as source A through the same aperture. 
Our flux calibration was based on an average of the values from
the spectra; the effects of the uncertainties will be discussed when
they are relevant to the analysis. In Figure~3 we present images of both the adjacent continuum and continuum-free H$\alpha$ line emission,
showing the positions of the slits used for both the optical
and near-infrared spectroscopy (Sections~2.4 and 2.5).

\begin{figure*}
\figurenum{3}
\plotfiddle{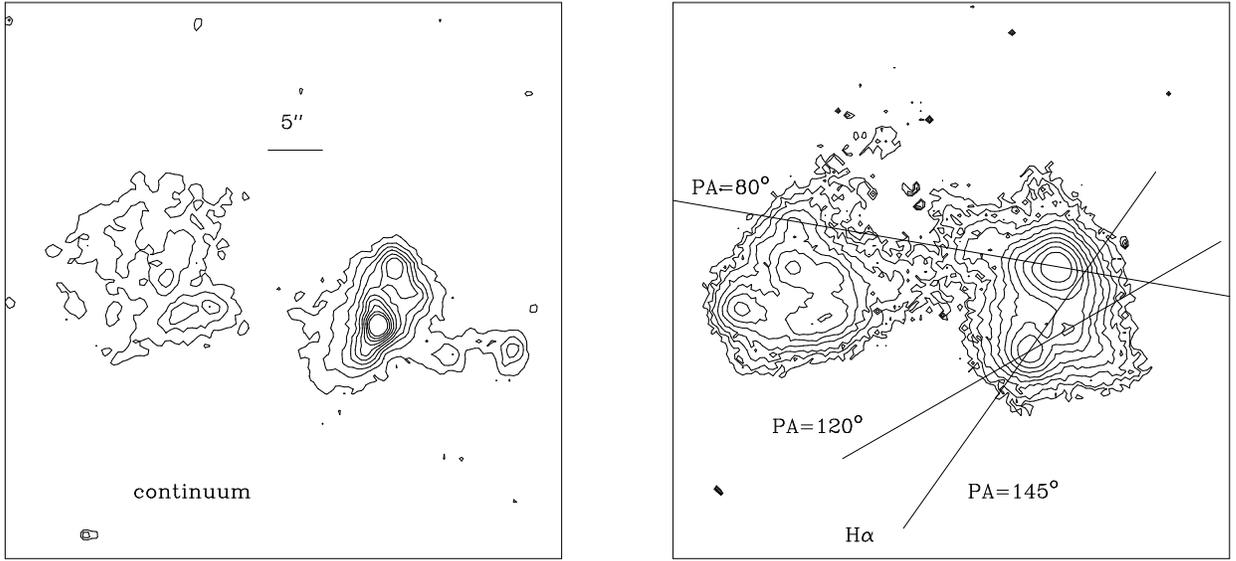}{425pt}{-90}{70}{70}{-260}{550}
\vspace{-7cm}
\caption{Contour plots on a linear scale of the continuum adjacent to
H$\alpha$ (left panel) and continuum free H$\alpha$ line emission
(right panel). The field of view ($51\arcsec \times 51\arcsec$) and
orientation  match those  of the NIC3 F166N image in presented in
Figure~1.}
\end{figure*}

\begin{deluxetable}{lcccccc}
\tablefontsize{\footnotesize}
\tablewidth{14cm}
\tablecaption{Optical Spectroscopy.}
\tablehead{\colhead{[O\,{\sc iii}]/H$\beta$} &
\colhead{[O\,{\sc i}]/H$\alpha$} &
\colhead{[N\,{\sc ii}]/H$\alpha$} &
\colhead{[S\,{\sc ii}]/H$\alpha$} &
\colhead{H$\alpha$/H$\beta$} &
\colhead{EW(H$\beta$)} &
\colhead{EW(H$\alpha$)}}
\startdata
\multicolumn{7}{c}{Source B1 + B16? \, $A_V = 4.2\,$mag} \\
\hline
2.59 & 0.055 & 0.20 & 0.27 & 12.4 & 15 & 170 \\
\hline
\multicolumn{7}{c}{Source B2 \, $A_V = 3.5\,$mag} \\
\hline
1.27 & 0.020 & 0.34 & 0.21 & 9.8 & 5 & 55 \\
\hline
\multicolumn{7}{c}{Source C \, $A_V = 2.1\,$mag} \\
\hline
1.42 & 0.018 & 0.35 & 0.17 & 6.0 & 93 & 607 \\
\hline
\multicolumn{7}{c}{Source C' \, $A_V = 2.1\,$mag} \\
\hline
1.16 & 0.042 & 0.31 & 0.29 & 6.0 & 45 & 293 \\
\hline
\multicolumn{7}{c}{Region north of A \, $A_V = 2.9\,$mag}\\
\hline
1.76 & 0.078 & 0.44 & 0.44 & 7.8 & 9 & 76 \\
\hline
\multicolumn{7}{c}{H\,{\sc ii} region 6.6\arcsec \, NW of B1 \,
$A_V = 2.9\,$mag}\\
\hline
1.57 & 0.013 & 0.34 & 0.19 & 7.6 & 48 & 300 \\
\hline
\multicolumn{7}{c}{H\,{\sc ii} region 7.8\arcsec \, NW of B1 \,
$A_V = 2.1\,$mag}\\
\hline
1.27 & 0.016 & 0.47 & 0.17 & 5.9 & 79 & 550 \\
\hline
\multicolumn{7}{c}{H\,{\sc ii} region 4.8\arcsec \, NW of B2 \,
$A_V = 1.7\,$mag}\\
\hline
1.16 & 0.016 & 0.34 & 0.21 & 5.2 & 45 & 277 \\
\hline
\multicolumn{7}{c}{H\,{\sc ii} region 2.4\arcsec \, NW of B2 \,
$A_V = 1.9\,$mag}\\
\hline
0.75 & 0.023 & 0.31 & 0.18 & 5.5 & 36 & 243 \\
\enddata
\end{deluxetable}

\subsection{$K$-band spectroscopy}
High ($\lambda/\Delta\lambda \simeq 3000$) and intermediate
($\lambda/\Delta\lambda \simeq 1000$)
resolution spectra centered at wavelengths $2.32\,\mu$m and
$2.20\,\mu$m respectively were obtained
at the Multiple Mirror Telescope with the FSPEC infrared spectrometer
(Williams et al. 1991) on February 11 and 12 1998. The slit size was 
$1.2\arcsec \times 30\arcsec$ with plate scale
0.4\arcsec pixel$^{-1}$. The slit was centered on B1 at position
angle PA $= 145\arcdeg$ to cover both sources B1 and B2, and at PA$=80\arcdeg$
centered on source C. The integration times were 96 minutes for sources
B1 and B2, and 28 minutes for source C for the high-resolution data, and
14 minutes for each of the three sources for the low-resolution data.

Observations were obtained for each galaxy at three or four positions along
the slit, integrating for 2 or 4 minutes at each position. Standard stars were
measured in a similar fashion (using shorter integrations), interspersed with
the galaxy observations. The standards were selected to be at similar air
masses as the galaxies so they could be used to correct for 
atmospheric absorptions. The data reduction process involves dark current
subtraction, flat-fielding, and sky subtraction. The correction for
atmospheric absorption is performed by dividing the galaxy spectrum
by the spectrum of a standard star observed at similar air mass.
The resulting spectrum is multiplied by a solar spectrum to
correct for the standard star absorption features. The wavelength
calibration used OH sky lines from the list in Oliva \& Origlia (1992).

We extracted spectra for B1, B2 and C with beam size $1.2\arcsec \times
1.2\arcsec$. Since conditions were non-photometric, the 
flux calibration was performed for each source by
extracting photometry with the same beam sizes on the NICMOS images taken
through filters NIC2 F212N and NIC2 F222M. The continuum fluxes at
$2.142\,\mu$m (for the observed wavelength of H$_2$) and
$2.188\,\mu$m (for the observed wavelength of
Br$\gamma$) were interpolated between these two filters. The line
fluxes were computed by multiplying the observed equivalent width of
the line by the continuum flux. Fluxes and equivalent widths for the
Br$\gamma$ and H$_2$(1-0)S(1) emission lines
are presented in Table~2. 

Figures~4a and 4b show the fully-reduced spectra. 
To check the NICMOS photometry,
we fitted the continuum slope between $2.22\,\mu$m and
$2.37\,\mu$m using the NIC2 F222M and NIC2 F237M fluxes obtained
for the corresponding apertures.  The fits are shown in these
figures as  dashed lines; the agreement is very good.
For the $^{12}{\rm CO}(2,0)$ band at
$2.293\,\mu$m we measure the spectroscopic index as defined in
Kleinmann \& Hall (1986), which is then transformed to a photometric
CO index. Measurements of both indices for B1, B2 and C are given
in Table~2.   Source C does not show CO bands. Our values
are in good agreement with Shier, Rieke, \& Rieke (1996) and 
Vanzi, Alonso-Herrero, \& Rieke (1998) for the components
in common with both works.

\begin{figure*}
\figurenum{4a}
\plotfiddle{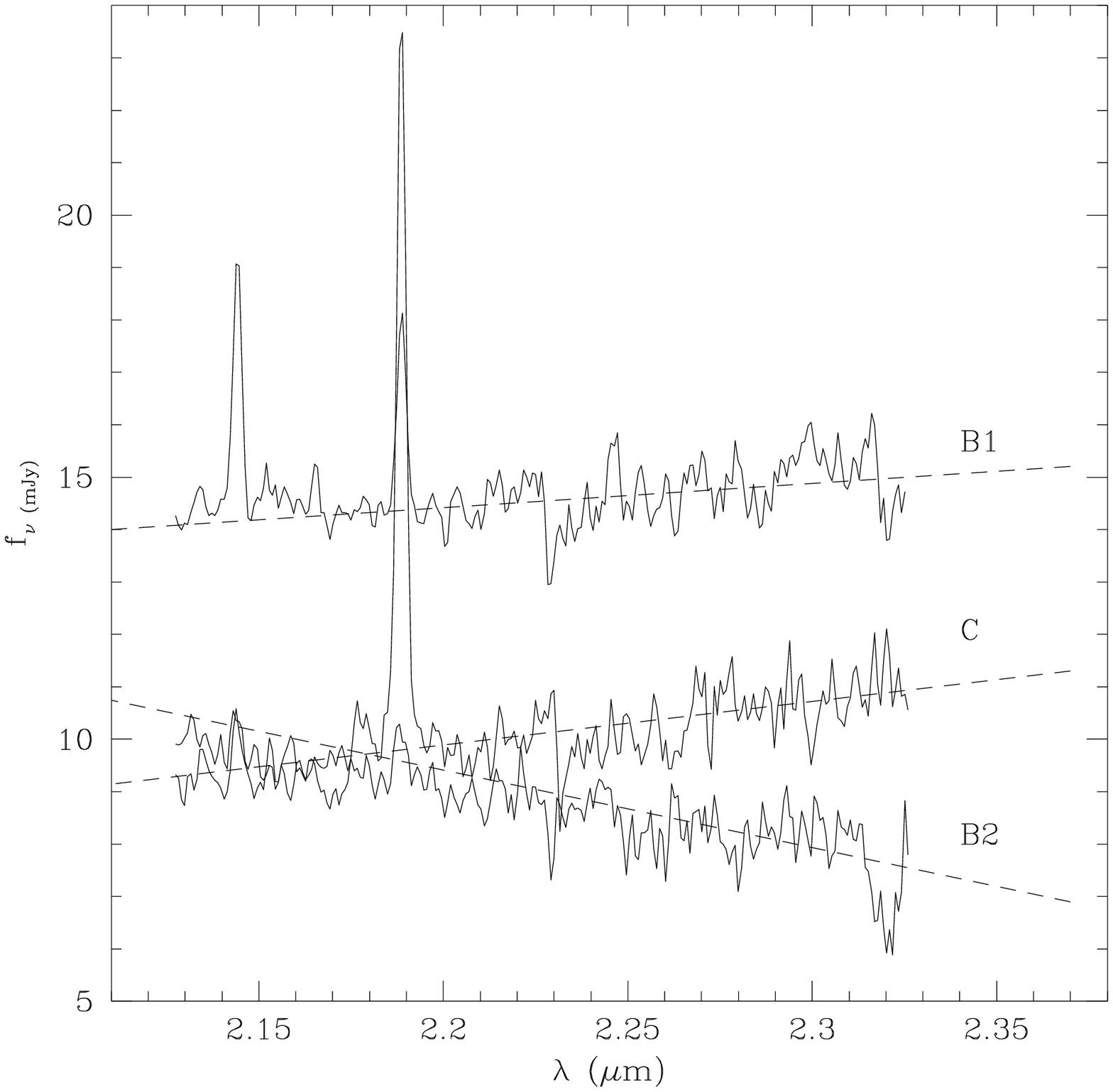}{425pt}{0}{50}{50}{-180}{50}
\vspace{-4.5cm}
\caption{Low resolution $K$-band spectroscopy for B1, B2 and C. The
dashed lines show the linear fits to the continuum between
$2.22\,\mu$m and $2.37\,\mu$m as measured through the NIC2 F222M and
NIC2 F237M filters.}
\end{figure*}

\begin{figure*}
\figurenum{4b}
\plotfiddle{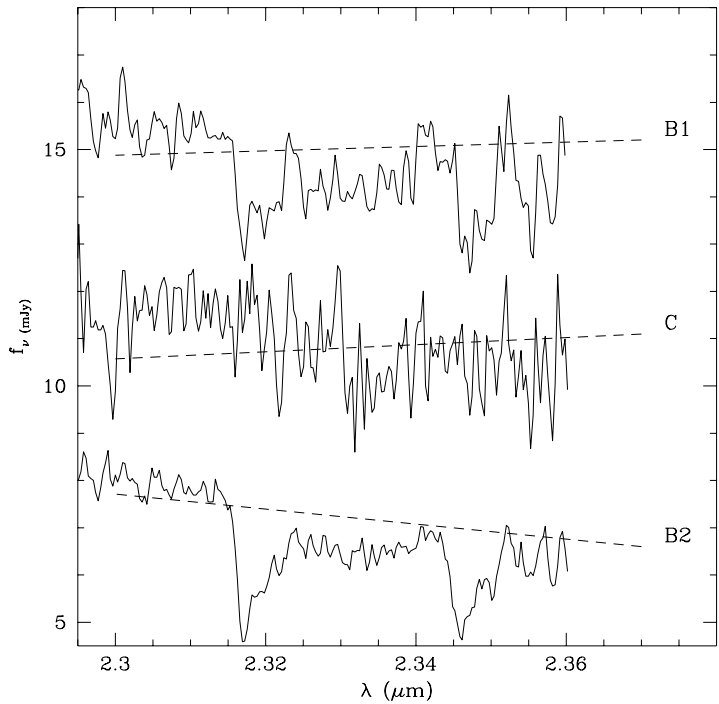}{425pt}{0}{50}{50}{-180}{50}
\vspace{-4.5cm}
\caption{Same as Figure~4a but for high resolution 
$K$-band spectroscopy.}
\end{figure*}

\begin{figure*}
\figurenum{5}
\plotfiddle{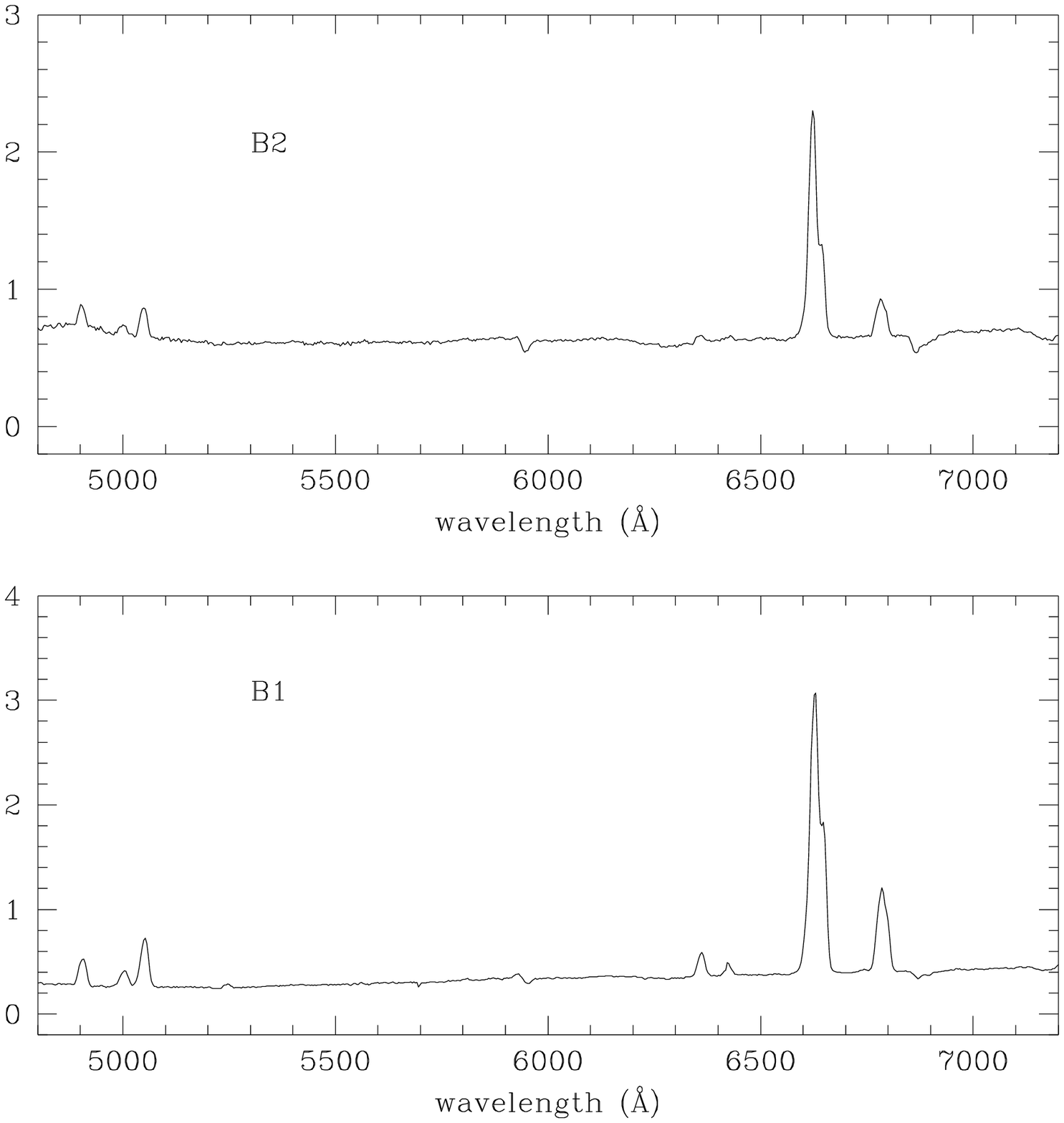}{425pt}{0}{40}{40}{-120}{150}
\plotfiddle{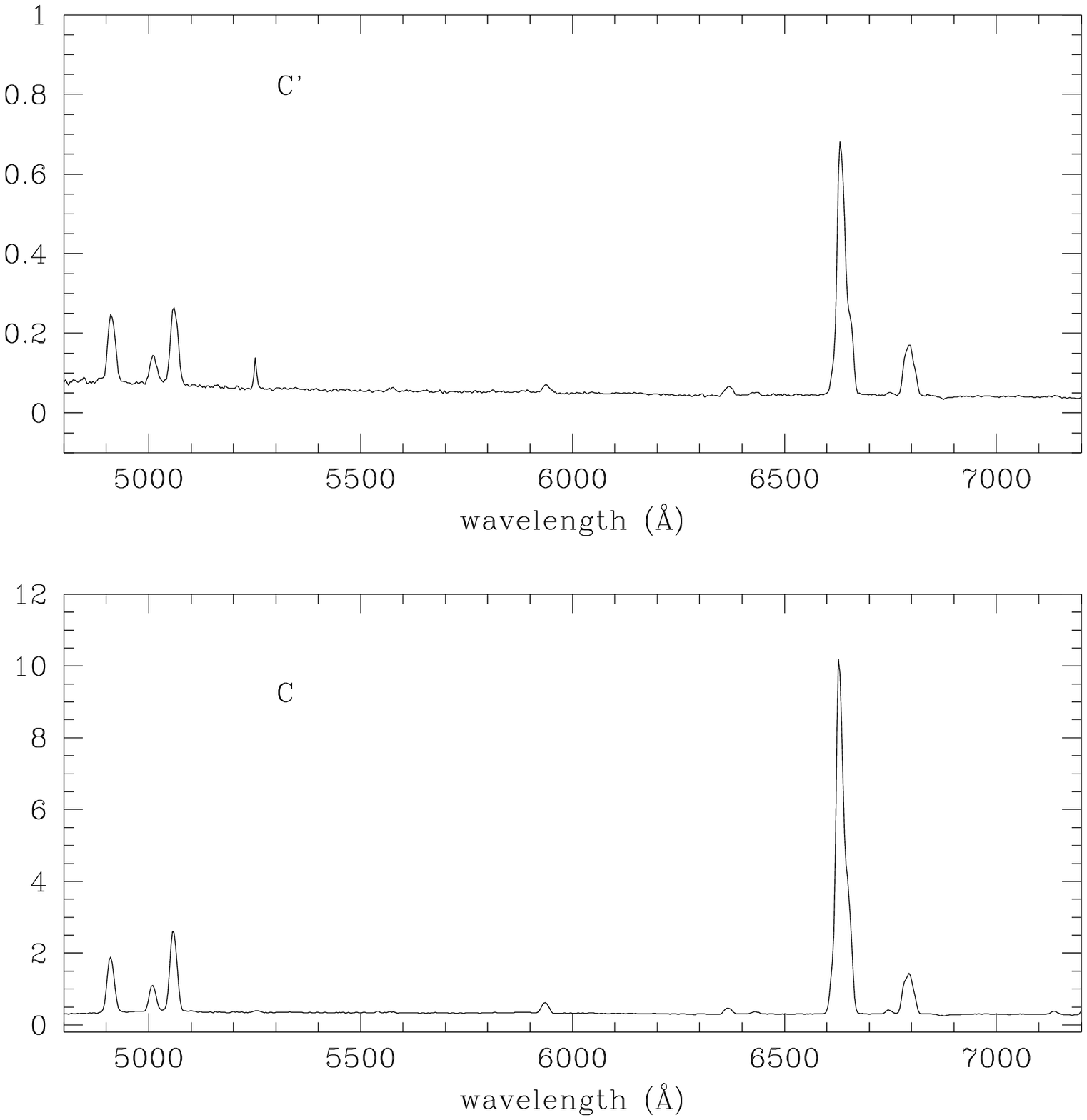}{425pt}{0}{40}{40}{-120}{350}
\vspace{-12.5cm}
\caption{Optical spectra (in arbitrary units) for B1, B2, C, and C',
along with a number of H\,{\sc ii} regions.}
\end{figure*}

\begin{figure*}
\figurenum{5}
\plotfiddle{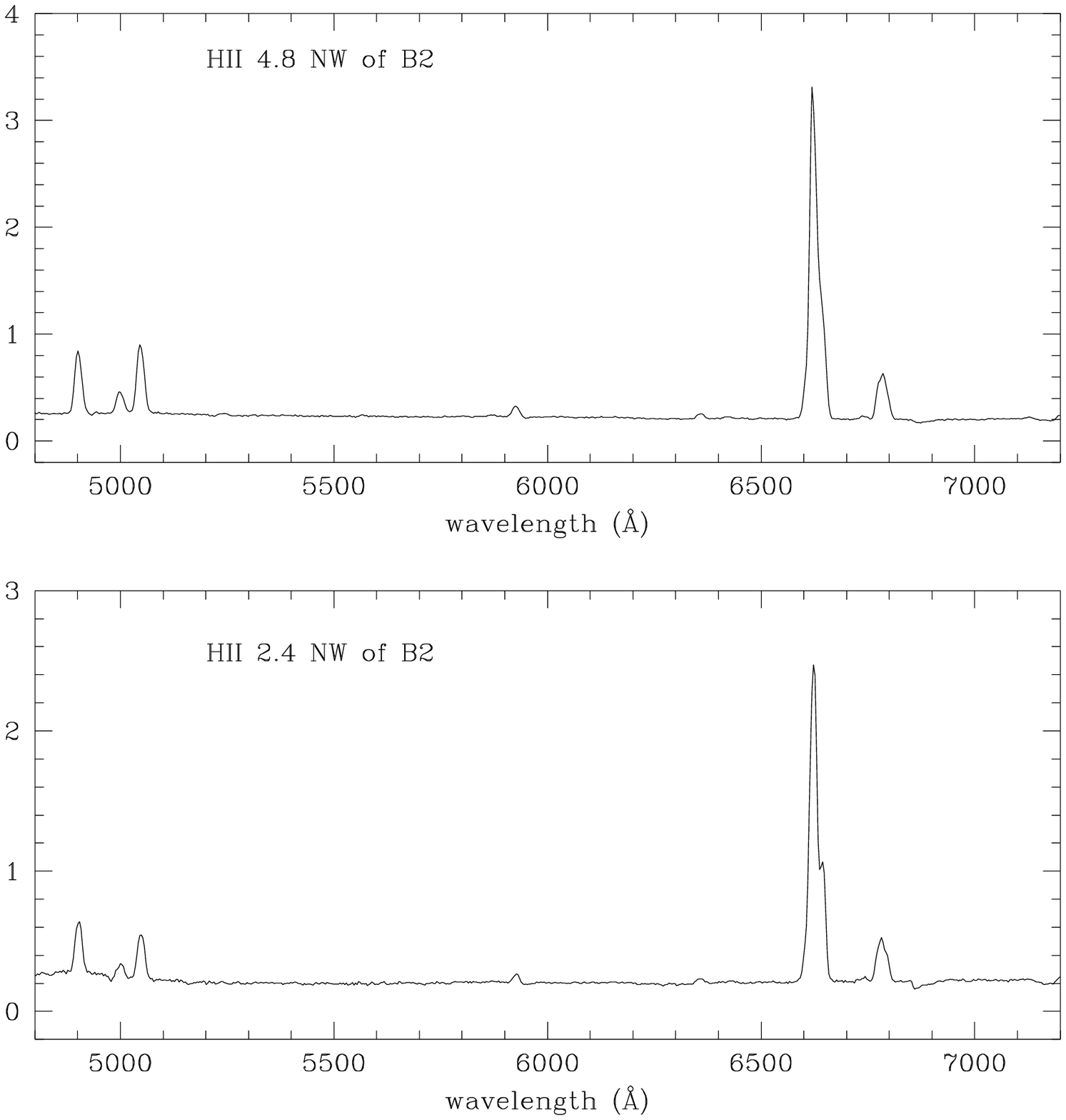}{425pt}{0}{40}{40}{-120}{150}
\plotfiddle{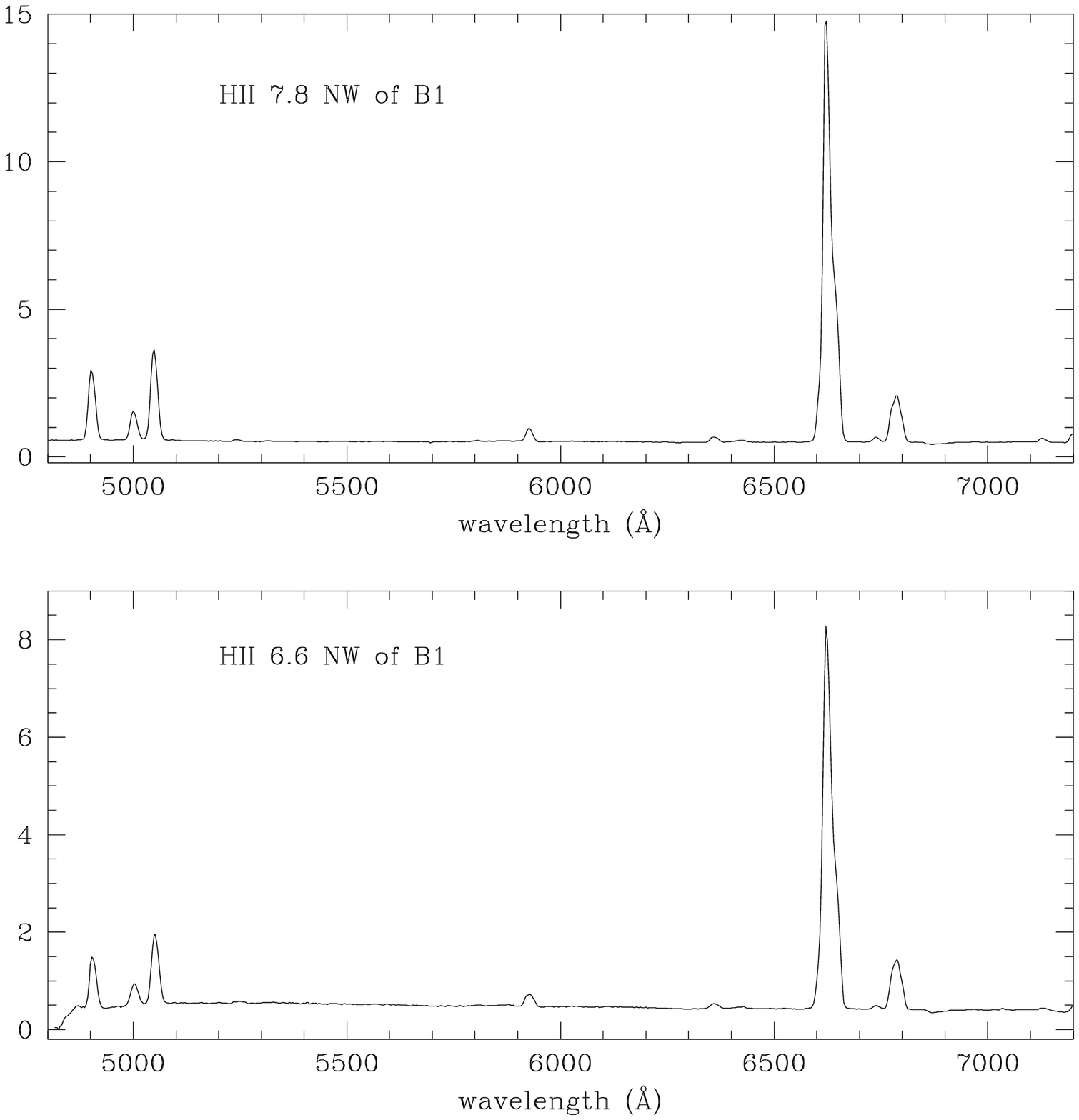}{425pt}{0}{40}{40}{-120}{350}
\vspace{-12.5cm}
\caption{Continued.}
\end{figure*}

\subsection{Optical Spectroscopy}
K. Luhman obtained  optical spectra of  NGC~3690 for us 
with the Red Channel Spectrograph at the Multiple Mirror Telescope on 
Mount Hopkins, on November 27--29  1997. He used the
270 g mm$^{-1}$ grating ($\lambda_{\rm blaze}=7300$\,\AA) to
measure spectra from 4600\,\AA \ to 8600\,\AA. A
$2\arcsec \times 180\arcsec$ slit with pixel size 0.6\arcsec \, 
pixel$^{-1}$ provided a spectral resolution of $\Delta\lambda = 
18\,$\AA. To derive the sensitivity function of the array he also 
observed the standard star BD+8 2015. The 2-D spectra were 
subtracted from bias-subtracted frames, corrected for the 
sensitivity function and wavelength calibrated using  He-Ar-Ne lamp 
spectra. We did not attempt to flux calibrate the data.

Spectra were extracted centered at B2 with position angles
 PA$=120\arcdeg$ and
PA$=145\arcdeg$, and centered near B1. Centering the slit on B1 with
the acquisition camera proved difficult because this nucleus is 
barely seen in the optical. Therefore, the coordinates obtained 
from the NICMOS images 
were used, and the slit was kept at the same  position angles
as for B2. Finally the slit was centered at C with position angle
PA$=80\arcdeg$. The position angles of the slits were chosen because 
PA$=145\arcdeg$ includes B1 and B2, and PA$=120\arcdeg$ 
includes the H\,{\sc ii} regions located northwest of B1.
The spectrum centered at B1 probably contains the source B16
which is located $\simeq 1.5\arcsec$ southwest of B1 (see Section~3.1).
The spectrum of B2 was extracted at the position of the peak of the continuum
near H$\alpha$. The spectrum centered at C at
PA = 80\arcdeg \, also contains C$^{\prime}$ and regions southwest of 
the nucleus of IC~694. 

For extracting the 1-D spectra we used 3 
pixel ($1.8\arcsec$) apertures. The positions of the slits and the regions for 
which we extracted spectra are shown in Figure~5. 
The fully reduced optical spectra  (not calibrated in flux)
are shown in Figures~5.a and 5.b for those regions for which we present line
ratios and equivalent widths in Table~3. The optical line ratios of all
the regions are consistent with those of H\,{\sc ii} as plotted in
diagnostic diagrams such those in Ho, Filippenko, \& Sargent al. (1993).
The visual extinctions were computed from the Balmer decrement.

\begin{figure*}
\figurenum{6}
\plotfiddle{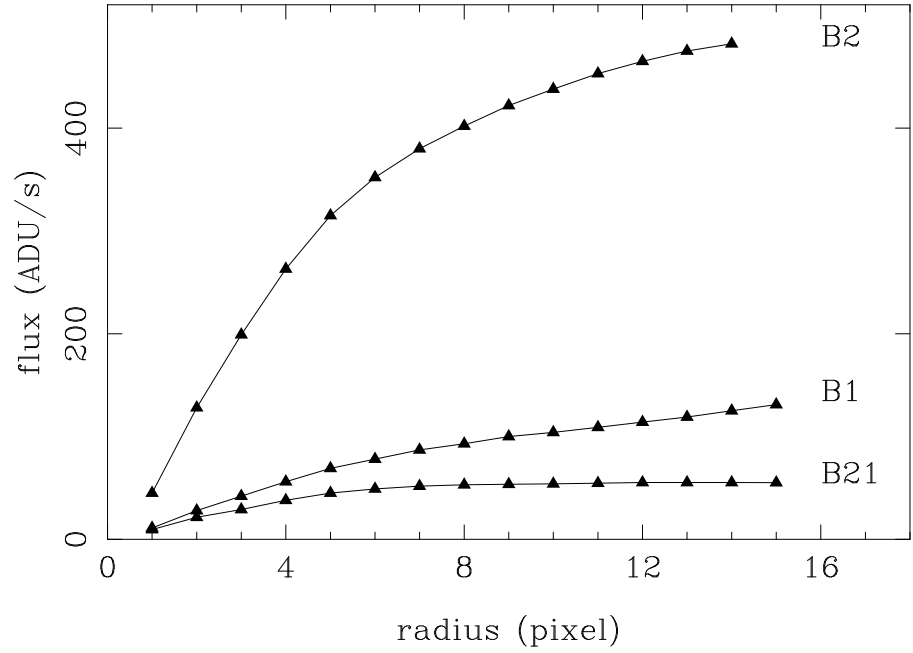}{425pt}{0}{70}{70}{-220}{150}
\vspace{-5.5cm}
\caption{Curves of growth for sources B1, B2 and B21 for the NIC1 F110M
filter.}
\end{figure*}

\section{RESULTS}
\subsection{Continuum Morphology}
Our NICMOS images of Arp~299 reveal an extremely complex morphology. 
Over the past few years it has been  claimed that 
the nucleus of IC~694 could harbor 
an obscured AGN. An important result of the NICMOS imaging is 
that no dominant point-like source is seen anywhere in the system 
that would be a logical AGN candidate. The NIC1 F110M image (Figure~2.1) 
of IC~694 shows that the nucleus is resolved into at least three
individual sources within a region of about 0.86\arcsec \, ($\simeq 175\,$pc) 
in diameter. These sources could be either young stellar clusters, 
or extinction shadows against a smooth source. Source B1 
(see Figure~7b for identifications) in NGC~3690 appears  to be 
extended at our resolution in the NIC1 F110M image, as does source 
B2. The curves of growth for sources B1, B2 and B21 for filter 
NIC1 F110M are shown in Figure~6 to illustrate that B2 is clearly extended,
whereas B1 appears marginally extended. As comparison we plot the curve of
growth of the source B21 which appears as a point source at our resolution. 
Sources B1 and B2 are also surrounded by a number of faint sources (see
Figure~7b for identifications and Table~4 for the relative distances
to either B1 or B2). Source C (which is seen as one source from
the ground) is now resolved into 3 sources that we will
refer to as C1, C2 and C3 (see Table~4 for relative positions). C1 and C2 are
extended whereas C3 seems to be point-like at the resolution of the
infrared images. The optical WFPC2 image, because of its higher
spatial  resolution, shows however that C3 is also extended.
Additional sources are located southwest of the C complex (i.e., C4 and C5)
and southeast of B1 (sources D1, D2, D3). The galaxy IC~694 also
shows faint sources, most of them located along the spiral arms (probably
H\,{\sc ii} regions), and labelled as A1, A2, A3, A4, A5 and A6
(Figure~7a). We can place an upper limit to the size of the unresolved 
sources from the NIC F110M image of $0.08\arcsec$ \ ($\simeq$ diffraction limit) or $\simeq 18\,$pc for the assumed distance. However, some of these 
faint sources appear to be slightly
resolved in the WFPC2 image.

A comparison between the NICMOS images
and the optical WFPC2 F606W image shows dramatically the effects of 
extinction. The latter image contains the H$\alpha$ emission line, so that low-extinction bright H\,{\sc ii} regions will show up in this filter. 
The most remarkable difference appears in the component IC~694. A 
careful alignment of the optical and near-infrared images to determine 
the position of the near-infrared peak on the optical image of IC~694 
shows that the source
(presumably the nucleus of the galaxy) is completely obscured in
the optical (see Figure~2.a). The spiral arm southwest of A is 
clearly detected in the optical image. The WFPC2 image of NGC~3690
shows a closer correspondence with the near-infrared image than for
IC~694, but still some differences are present. The very high spatial
resolution of the WFPC2 image shows that both sources B1
and B2 are detected and extended (note the morphology of the 
ground-based continuum image near H$\alpha$ where the brightest 
source is B2, Figure~3),
although B2 is much brighter than B1 at optical wavelengths (as already
known from ground-based  observations). Most of the faint sources seen
in the NICMOS images (i.e., the B's, C's and D's sources) are
also seen in the optical image, however some of them must be very obscured
because they are very faint in the optical. On the other hand, source
B16, which is barely seen in the NIC2 F160W image, appears to be almost
as bright as B1 in the optical.

\begin{figure*}
\figurenum{7a}
\plotfiddle{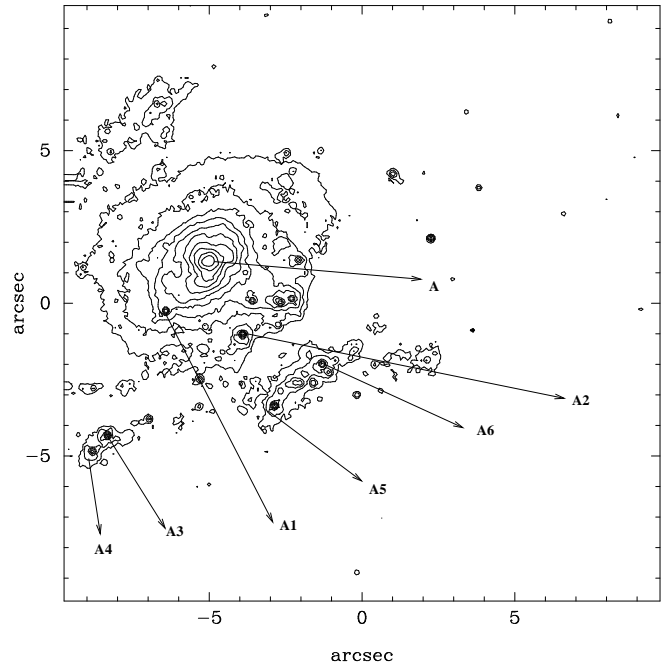}{425pt}{-90}{50}{50}{-200}{450}
\vspace{-5.5cm}
\caption{Contour plot on a linear scale of the NIC2 F160W image
of IC~694. We mark the positions of the A sources (Table~4). Orientation
is north up, east to the left. The field of view is
$19.5\arcsec \times 19.5\arcsec$.}
\end{figure*}

\begin{figure*}
\figurenum{7b}
\plotfiddle{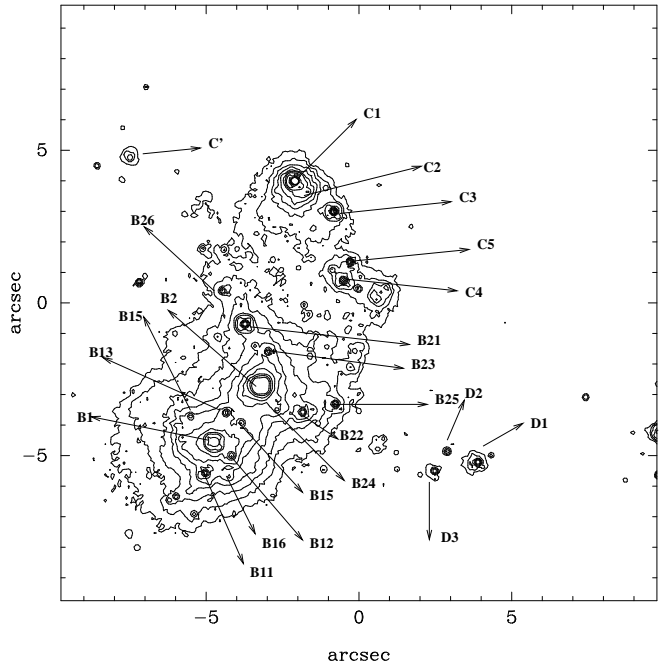}{425pt}{-90}{50}{50}{-200}{450}
\vspace{-5.5cm}
\caption{Contour plot on a linear scale of the NIC2 F160W image
of NGC~3690. We mark the positions of the B, C and D sources
(Table~4). Orientation and field of view as in Figure~7a.}
\end{figure*}

\begin{deluxetable}{lccccccc}
\tablefontsize{\footnotesize}
\tablewidth{18cm}
\tablecaption{Photometry of A, B1, B2, C1, and faint sources.
Relative distances and fluxes.}
\tablehead{\colhead{Source} & \colhead{$\Delta$RA} &
\colhead{$\Delta$Dec} & \colhead{$f($F110M)} & \colhead{$f($F160W)} &
\colhead{$f($F222M)} & \colhead{$f($F237M)} &
\colhead{$f$(F606W)}\\
\colhead{} & \colhead{s} & \colhead{\arcsec} & \colhead{mJy} &
\colhead{mJy} & \colhead{mJy} & \colhead{mJy} &  \colhead{erg cm$^{-2}$
s$^{-1}$ \AA$^{-1}$} }

\startdata
A   & 0 & 0 &
0.886&   3.504&   7.280&   7.458& \nodata \nl
A1  & 0.17 & -1.62 &
0.036&   0.100&   0.159&   0.121& \nodata \nl
A2  &   -0.14 & -2.39   &
0.057&   0.124&   0.129&   0.097 & \nodata \nl
A3  &   0.41 & -5.70 &
0.047&   0.086&   0.097 &   0.075 & \nodata \nl
A4  &   0.48 & -6.20 &
0.023&   0.063&   0.073&   0.058& \nodata \nl
A5  &   -0.27 & -4.71 &
0.061&   0.098&   0.108&   0.085& \nodata \nl
A6  &   & &
0.050&   0.079&   0.080&   0.057& \nodata \nl
\nl
B1  &   0. & 0 &
0.824&   2.909& 10.739&  13.827& $0.19\times 10^{-15}$  \nl
B11 &   0.04 & -1.00 &
0.158&   0.325&   0.362&   0.333& $0.24\times 10^{-16}$ \nl
B12 &   -0.08 & -0.42 &
0.072&   0.279&   0.372& \nodata& \nodata \nl
B13 &   -0.05 & 0.91 &
0.087&   0.254&   0.397&   0.298& \nodata \nl
B14 &   -0.11 & 0.64 &
0.060&   0.104& \nodata& \nodata& \nodata \nl
B15 &   0.09 & 0.83 &
0.044&   0.115&   0.138& \nodata& \nodata \nl
B16 &   -0.06 & -1.10 &
0.086&   0.061& \nodata& \nodata& $0.10\times 10^{-15}$ \nl
\nl
B2  &   0 & 0 &
3.678&   5.737&   6.676&   5.525& $0.89\times 10^{-15}$ \nl
B21 &   0.07 & 2.01 &
0.440&   0.787&   0.806&   0.663& $0.23\times 10^{-15}$ \nl
B22 &   -0.17 & -0.85 &
0.123&   0.274&   0.338&   0.252& $0.90\times 10^{-17}$ \nl
B23 &   -0.03 & 1.14 &
0.088&   0.173&   0.152&   0.127& $0.18\times 10^{-16}$ \nl
B24 &   -0.07 & -0.77 &
0.054&   0.091&   0.093& \nodata& \nodata \nl
B25 &   -0.30 & -0.60 &
0.042&   0.096&   0.104& \nodata& $0.16\times 10^{-16}$ \nl
B26 &   0.16 & 3.10 &
0.039&   0.067&   0.060& \nodata& $0.20\times 10^{-16}$ \nl
\nl
C1  &   0 & 0 &
0.199&   1.460&   8.4071&  11.152& $0.72\times 10^{-16}$ \nl
C2  &   -0.05 & -0.31 &
0.171&   0.255& \nodata& \nodata& $0.22\times 10^{-15}$ \nl
C3  &   -0.16 & -1.00 &
0.199&   0.307&   0.307&   0.206& $0.76\times 10^{-16}$ \nl
C4  &   -0.20 & -3.25 &
0.119&   0.224&   0.208&   0.154& $0.51\times 10^{-16}$ \nl
C5  &   -0.23 & -2.63 &
0.051&   0.084&   0.083& \nodata& $0.32\times 10^{-16}$ \nl
C'   & 0.68 & 0.77 & \nodata & 0.081 & 0.165 & 0.160 & \nodata \nl
\nl
D1  &   -1.09 & -0.66 &
0.087&   0.136&   0.134&   0.087& $0.47\times 10^{-16}$ \nl
D2  &   -0.96 & -0.31 &
0.013&   0.030&   0.027& \nodata& \nodata \nl
D3  &   -0.91 & -0.94 &
0.031&   0.048&   0.046& \nodata& $0.10\times 10^{-16}$ \nl
\enddata

\tablecomments{Relative distances for A1 through A6 are from A,
relative distances for B11 through B16, and D1 through D3 are from B1,
relative distances for B21 through B25 are from B2 and relative distances
for C2 through C5 are from C1. The photometry for A, B1 and B2
is through a 0.9\arcsec-diameter photometry, whereas that of
the faint sources is for a 0.5\arcsec-diameter aperture (corrected
for aperture effects). Errors for the measured fluxes are 
7\% for the bright sources and 10\% for the faint sources.}
\end{deluxetable}

\begin{deluxetable}{lccccc}
\tablefontsize{\footnotesize}
\tablewidth{14cm}
\tablecaption{Observed $H$-band magnitudes, colors and (CO)$_{\rm NICMOS}$ 
index. }
\tablehead{\colhead{Source} & \colhead{$H$} & 
\colhead{$J-H$} & \colhead{$H-K$} &
\colhead{CO} & \colhead{$R-J$}}

\startdata
A	&13.724	&1.330	&1.361	&$-0.025$ & \nodata \nl
A1	&17.588	&1.073	&1.039	&0.239 &\nodata \nl
A2	&17.357	&0.908	&0.570	&0.249 &\nodata \nl
A3	&17.746	&0.807	&0.649	&0.229 &\nodata \nl
A4	&18.095	&1.049	&0.686	&0.203 &\nodata \nl
A5	&17.614	&0.720	&0.632	&0.209 &\nodata \nl
A6	&17.846	&0.701	&0.539	&0.256 &\nodata \nl
\nl
B1	&13.927	&1.230	&2.061	&$-0.288$ & 2.18 \nl
B11	&16.305	&0.880	&0.638	&0.079  & 2.38 \nl
B12	&16.473	&1.291	&0.837	&\nodata &\nodata \nl 
B13	&16.575	&1.108	&1.019	&0.249  &\nodata \nl
B14	&17.539	&0.770	&\nodata &\nodata & \nodata \nl	 
B15	&17.430	&1.034	&0.720	&\nodata &\nodata \nl
B16	&18.121	&0.190	&\nodata &\nodata & $-0.15$	 \nl
\nl
B2	&13.190	&0.699	&0.687	&0.172 & 1.84 \nl
B21	&15.347	&0.788	&0.548	&0.178 & 1.15 \nl
B22	&16.493	&0.930	&0.751	&0.252 & 2.97 \nl
B23	&16.991	&0.853	&0.386	&0.165 & 2.09 \nl
B24	&17.687	&0.750	&0.541	&\nodata &\nodata \nl
B25	&17.636	&0.939	&0.618	&\nodata & 1.57\nl
B26	&18.028	&0.766	&0.418	&\nodata & 1.08\nl
\nl
C1	&14.676	&1.705	&2.646	&$-0.327$ & 1.97 \nl
C2	&16.570	&0.672	&\nodata &\nodata &0.13 \nl	 
C3	&16.368	&0.694	&0.522	&0.331 & 1.37 \nl
C4	&16.712	&0.823	&0.443	&0.259 & 1.34 \nl
C5	&17.781	&0.727	&0.510	&\nodata& 0.89 \nl
\nl
D1	&17.253	&0.703	&0.505	&0.352 & 1.04\nl
D2	&18.877	&0.949	&0.392	&\nodata & \nodata \nl
D3	&18.372	&0.699	&0.462	&\nodata & 1.55 \nl

\enddata
\tablecomments{Errors in the colors are $\pm0.10$ and $\pm 0.14$
for the bright and faint sources respectively, whereas the errors
associated with the (CO)$_{\rm NICMOS}$ are of the order of 20\%. }
\end{deluxetable}

The effects of extinction are even more dramatic when the NICMOS images
are compared with the ultraviolet FOC F220W image. We carefully 
cross-correlated the UV sources reported by Meurer et al. (1995, see their
table~10), with the infrared ones, using relative coordinates.
The brightest source in the UV image (NGC3690-1 in their notation) 
is identified with our source B21. The UV source NGC~3690-2 corresponds 
to B2. The double source NGC~3690-5 and NGC~3690-8 is located at 
the position of source B16, and therefore does not correspond 
to source B1 (as Meurer et al. 1995 indicated). Source B1 appears 
totally obscured at ultraviolet wavelengths. The positions of 
sources NGC~3690-6, NGC~3690-7, NGC~3690-9 and NGC~3690-10 are approximately
consistent with those of C2, C3, C5 and C4 in our notation. Also 
noticeable is that the D complex is relatively bright in the UV, although 
it does not stand out in the IR. All the UV sources detected in
 NGC~3690 are embedded in a UV background which extends to greater
radii than the individual sources. IC~694 is very faint in the
ultraviolet image (not shown here) with only the southeast spiral arm
and some very diffuse and faint emission near A.

These comparisons are a warning about the use of the UV to study starbursts,
particularly ones that are less well resolved than Arp~299. The UV light
can be dominated by lightly obscured regions that represent only a fraction
of the true starburst region and which may be atypical of the integrated
properties (e.g., the extended UV backgrounds in NGC~3690 and the
relative suppression of IC~694).

\subsection{Emission Line Morphology}
The [Fe\,{\sc ii}]$1.644\,\mu$m and the Pa$\alpha$ continuum-subtracted
images show bright emission from A, B1, and C  (Figures~2.a and 2.b). From
both these images it appears that very little line emission originates
from source B2, confirming the result of Fischer, Smith \& Glaccum (1991). 
The [Fe\,{\sc ii}]$1.644\,\mu$m and Pa$\alpha$  images trace the 
emission from H\,{\sc ii} regions located in the spiral arms SE of 
IC~694. The Pa$\alpha$ emission of the nucleus of IC~694 is similar 
to the radio H92$\alpha$ morphology (Zhao et al. 1997). However, the 
SE emission seen in Zhao et al. (1997) map does not correlate exactly 
with the Pa$\alpha$ morphology, possibly due to some obscuration effects
at the infrared wavelengths. There are a number of H\,{\sc ii} 
regions northwest of B1, and east of C reaching all the way to 
source C$^\prime$ and extending toward IC~694. In the Pa$\alpha$ 
image we can also resolve the B16 source.

The H$_2$ image of IC~694 shows intense emission
from the nucleus of the galaxy together with a beautiful butterfly-like
diffuse emission extending over approximately 3 arcsec (see Figure~2.a).
The molecular hydrogen image of NGC~3690 shows mainly point-like emission
from sources B1, B2 and C, and some diffuse emission from source C$^\prime$
(Figure~2.b).

\subsection{Photometry of A, B1, B2, C, and the faint sources}

To study the faint sources in Arp~299, we need accurate magnitudes 
and colors. However, some of these sources, such as the faint objects 
near B1 and B2,  are located on a local background with  a substantial 
gradient. To subtract the contribution of 
the underlying galaxy, first we extracted subimages of each source, then 
we produced a two dimensional model of
the local background for each source. The source plus background fluxes were
fitted unidimensionally along both rows and columns. The source 
was represented with two Gaussians with different FWHMs and amplitudes, 
whereas we used a linear fit to the local background. Although the NICMOS PSF
is not perfectly represented by Gaussian functions, the one-dimensional
emission profiles were found to be fitted well. The two fits along rows 
and columns were then averaged together to produce independent 2-D 
source and the background models. The model background was 
subsequently subtracted from the original sub-image.

\begin{deluxetable}{lcccc}
\tablefontsize{\footnotesize}
\tablewidth{12cm}
\tablecaption{Extinction to the gas for the different components of
Arp~299.}
\tablehead{\colhead{Aperture} & \colhead{source} & \colhead{$f$(Pa$\alpha$)} &
\colhead{$f$(Pa$\alpha)/f$(H$\alpha)$} & \colhead{$A_V$(screen)}\nl
& & \colhead{erg cm$^{-2}$ s$^{-1}$} & & \colhead{mag}}
\startdata
circular $5\arcsec$ & A  & $3.7\times 10^{-13}$ & 1.9 & 5 \\
                & B1 & $2.3\times 10^{-13}$ & 0.5 & 2 \\
                & C  & $4.4\times 10^{-13}$ & 0.4 & 2 \\
\hline
$4.5\arcsec\times 2\arcsec$ & A & $3.1 \times 10^{-13}$ & $1.5-3.0$
& $4-5$ \nl
                            &B1 & $1.7 \times 10^{-13}$ & $0.5-1.0$
& $3-4$ \nl
                            & C & $3.3 \times 10^{-13}$ & $0.4-0.8$
& $2-3$ \nl
\hline
$2\arcsec\times 2\arcsec$   & A & $2.4 \times 10^{-13}$ & $3-5$
& $5-6$ \nl
                            &B1 & $1.3 \times 10^{-13}$ & $0.7-0.9$
& $3-4$ \nl
                            & C & $2.7 \times 10^{-13}$ & $0.6-0.9$ &
$3-4$ \nl
\hline
\hline
{Aperture} &{source} & \colhead{$f$(Pa$\alpha$)} &
{$f$(Br$\gamma)/f$(Pa$\alpha)$} & {$A_V$(screen)}\nl
& & erg cm$^{-2}$ s$^{-1}$ & &mag \nl
\hline
$2.4\arcsec\times 4.7\arcsec$\tablenotemark{a}   & A  & $3.5\times 10^{-13}$
& 0.115   & $24\pm 7$ \nl
$1.2\arcsec\times 1.2\arcsec$   & B1 & $8.3\times 10^{-14}$ & 0.102 &
$16\pm 7$  \nl
                                & C  & $1.8\times 10^{-13}$ & 0.142  &
$37\pm 10$ \nl
$1.1\arcsec\times 2.2\arcsec$\tablenotemark{a}   & B1 & $9.6\times 10^{-14}$
& 0.104  &  $17\pm7$ \nl
\hline
\hline
{Aperture} &{source} & \colhead{$f$(Pa$\alpha$)} &
{$f$(Pa$\alpha)/f$(Pa$\beta)$} & {$A_V$(screen)}\nl
& & erg cm$^{-2}$ s$^{-1}$ & & mag \nl
\hline
$2.4\arcsec\times 4.7\arcsec$\tablenotemark{b}   & A  & $3.5\times 10^{-13}$
& 5.7   & $9\pm 2$      \nl

\enddata
\tablenotetext{a}{Fluxes of Br$\gamma$ calculated from EW of Br$\gamma$ and
NICMOS continuum calibration (Craig Kulesa private communication).}
\tablenotetext{b}{Flux of Pa$\beta$ calculated from EW of Pa$\beta$
and NICMOS continuum calibration (Craig Kulesa private communication).}
\end{deluxetable}

Aperture photometry was done on the background subtracted images.
Most of the faint sources appear to be unresolved, so the photometry 
was obtained through relatively small apertures (typically 0.5\arcsec 
\ in diameter). The final magnitudes and fluxes were corrected for 
the behavior of the PSF so that the measurement included all the 
flux associated with 
a point source. The errors in the photometry for each individual 
source were computed as the
standard deviation of the subimage resulting from the subtraction of
the background+source model image from the original image\footnote{
$\sigma_{\rm b}^2 = \sum \frac{({\rm residual(i,j)} - <{\rm residual}>)^2}
{n(n-1)}$ with $n = i\times j$ and residual = image - model}. These errors
are approximately $0.1\,$mag for the faintest sources, decreasing to 
$0.07\,$mag for bright ones.

Fluxes (in mJy) measured through the NIC1 F110M, NIC2 F160W, NIC2 F222M and
NIC2 F237M filters along with the fluxes (in erg cm$^{-2}$ s$^{-1}$) through the 
WFPC2 F606W filter are given in Table~4. For the WFPC2 F606W filter we used the 
photometric calibration from the header of the image, since this filter was 
not included in the set of WFPC2 calibrated filters.
In Table~5 we give the $H$-band magnitudes and 
$J-H$, $H-K$ and $R-J$ colors. The  $J-H$ and $H-K$
colors were obtained by convolving an empirical galaxy spectral energy
distribution with both the NICMOS and $J,H,K$ filter transmission
functions and detector response functions. The same convolutions were also
computed for the galaxy reddened by as much as 40 magnitudes. The
convolutions were compared to the values for the standard star 
P330E in each filter set to
derive a transformation between the NICMOS F110M-F160W-F222M system and
the $JHK$ system. This procedure also corrected for the large color
difference between the standard star and our objects in the rather
broad F160W filter. The Johnson $R$ magnitudes are interpolated 
between the fluxes  at 6060\,\AA \ and at $1.10\,\mu$m. 

In addition we measure a photometric NICMOS CO index defined as,

\begin{equation}
{\rm CO}_{\rm NICMOS} = \frac{f({\rm F222M}) - f({\rm F237M})}{f({\rm F222M})}
\end{equation}

\noindent This quantity does  not formally correspond with the standard
photometric CO index, but as Figure~8 illustrates, it is nearly
equal to the photometric CO index. The relation plotted in Figure~8 was
derived by integrating the SED for an old stellar population across both the
two NICMOS filter functions and the ground-based filter functions. 
The SED was redshifted to .010 and also reddened by amounts ranging from 
$A_V=0\,$mag to $A_V=40\,$mag. The measured values for both the bright and 
faint sources are given in Table~5. The NICMOS CO index is subject 
to the same caveats as the ground-based index because it measures the 
slope of the continuum between $2.22\,\mu$m and $2.37\,\mu$m as well 
as the strength of the CO absorption. As we will discuss in 
Section~3.3.2 it is sensitive to both extinction effects and 
dilution of the CO bands by hot dust emission.

\begin{figure*}
\figurenum{8}
\plotfiddle{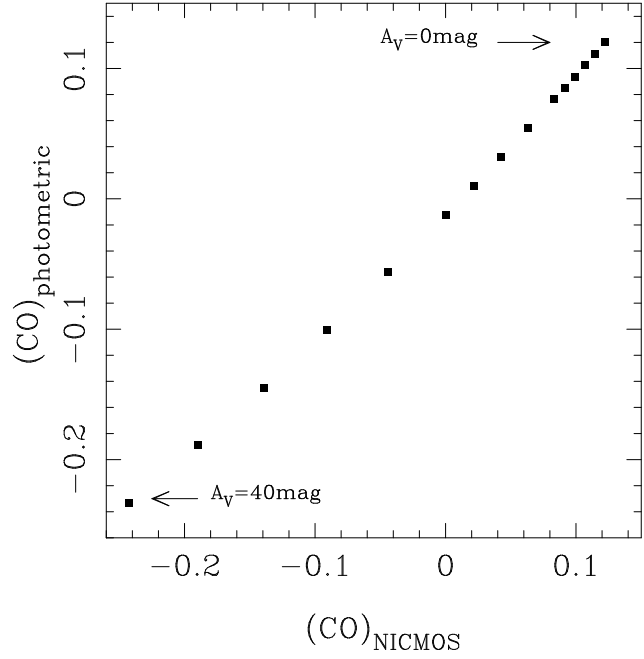}{425pt}{0}{70}{70}{-150}{150}
\vspace{-5.5cm}
\caption{Relation between the photometric (CO) index and   
the (CO)$_{\rm NICMOS}$ 
for an elliptical galaxy redshifted to $z=0.01$ and different values
of the extinction from $A_V=0\,$mag through $A_V=40\,$mag.}
\end{figure*}



\begin{deluxetable}{lcccc}
\tablefontsize{\footnotesize}
\tablewidth{11cm}
\tablecaption{Extinction to the stars, corrected colors and NICMOS CO indices.
Dust emission.}
\tablehead{\colhead{Source} & \colhead{$A_V$} &
\colhead{$(H-K)_{\rm corr}$} &
\colhead{CO$_{\rm NICMOS}$(corrected)} &
\colhead{($f_{\rm dust}/f_{\rm total})_K$}\\
\colhead{(1)} &\colhead{(2)} & \colhead{(3)} & \colhead{(4)} &
\colhead{(5)}}
\startdata
A  & 6.5 & 1.01 & 0.03 & \nodata \nl
B1 & 5.2 & 1.78 & $-0.24$ &  \nodata \nl
C1 & 9.9 & 1.90 & $-0.23$ & \nodata \nl
\hline
A  & $24\pm 7$  & $0.08\pm0.40$  & $0.15\pm0.04$ & 0 \nl
B1 & $17\pm 7$  & $1.20\pm0.40$  & $-0.15\pm0.04$ & $\simeq 0.6$  \nl
B2 & $3.5\pm0.5$ & $0.31\pm0.03$ & $0.19\pm0.01$ & 0 \nl
C1 & $37\pm 10$  & $0.43\pm 0.70$ & $0.00\pm 0.06$  & $<0.2$ \nl
\enddata
\tablecomments{The errors  in columns~(3) and (4) take into account 
the uncertainties associated with the extinction.}
\end{deluxetable}

\subsection{Extinction}

To interpret the observations requires an accurate estimate of the 
extinction 
to each source. The extinction to the gas can be estimated from hydrogen 
recombination line ratios, so long as the comparison is based on 
measurements with the same beam sizes. To determine the extinction 
to the stars one can compare observed colors with the colors of an 
evolved stellar population. The infrared colors of evolved stellar 
populations are nearly universal, independent of age. In addition a 
correction for the dust contribution to the $K$-band fluxes -if any- 
is required prior to dereddening the stellar colors. 

The largest uncertainty in the extinction by far comes from the unknown 
distribution of dust within the source. Meurer et al. (1995) from 
their study of 
the UV properties of starburst galaxies concluded that the dust 
geometry has to 
be in a foreground screen configuration near the starburst. However, 
UV-selected
galaxies will be biased toward low extinction and possibly an 
unrepresentative distribution of the dust. In addition, as emphasized in 
Section~3.1, the UV may be dominated by non-representative regions within
a galaxy. Furthermore, within this study NGC~3690 
was a notable exception because it appeared to have
patchy extinction. Other authors (e.g., McLeod et al. 1993 for M82; 
Genzel et al. 1995 for NGC~7469; and Sugai et al. 1999 for Arp 299) 
find that a simple foreground screen model does not provide a good 
fit for relatively dust embedded starbursts. They used  a model in 
which dust and gas are homogeneously mixed, and showed that the 
foreground screen model gives a lower limit -- sometimes by a 
large factor -- to the extinction.

\subsubsection{Extinction to the gas}
We use a foreground dust screen model and consider its effects on the 
hydrogen recombination lines. For the infrared extinction law we
use a polynomial fit to the $A_J$, $A_H$ and $A_K$ values
given in He et al. (1995). The resulting relative extinctions are,
$A$(Pa$\beta) = 0.254\times A_V$,
$A$(Pa$\alpha) = 0.128\times A_V$,
$A$(Br$\gamma) = 0.114\times A_V$. For the optical lines we use
$A$(H$\beta) = 1.20 \times A_V$ and $A$(H$\alpha) = 0.73 \times A_V$
(Rieke \& Lebofsky 1985). The intrinsic values for the Case B hydrogen line
ratios are taken from Hummer \& Storey (1987) for physical
conditions $T_{\rm e} = 10,000\,$K and $n_{\rm e} = 100\,$ cm$^{-3}$.

The results for different line ratios and aperture sizes are presented
in Table~6. In this table the fluxes of the hydrogen recombination lines
are ratioed to the Pa$\alpha$ flux, which was measured through the 
corresponding aperture from our NICMOS images. The two values of the Pa$\alpha$ to H$\alpha$ ratios and extinctions for the $4.5\arcsec \times 2\arcsec$ and
$2\arcsec \times 2\arcsec$  apertures correspond to the uncertainty 
of the H$\alpha$ flux calibration, as discussed in Section~2.3.
The results from the Br$\gamma$ to Pa$\alpha$ line ratios are
presented in the second part of this table. We did not obtain
$K$-band spectroscopy for source A, but we made use of the data 
in Kulesa et al. (1999). Instead of using their flux calibration, 
we measured the continuum fluxes from the NICMOS images and 
used their equivalent widths of Br$\gamma$ and
Pa$\beta$, to compute the line fluxes. The errors listed in Table~6 for the
$A_V$ derived from the Br$\gamma$ to Pa$\alpha$ and
Pa$\alpha$ to Pa$\beta$ line ratios are simply the propagation of the errors
in the measured equivalent widths. In Table~3 we give the extinctions
derived from the Balmer decrement. 

From Table~6 the extinction tends to be higher when line ratios involving 
longer wavelengths are employed. For instance, the extinction towards 
source A increases from $A_V = 5.8\,$mag from the Balmer decrement up to $A_V = 24\pm7\,$mag for ratios involving longer wavelengths. A similar trend is found for source C (C1+C2) where extinctions range from $A_V = 2.1\,$mag from 
the optical line ratios (see Table~3) to $A_V= 37\pm10\,$mag from the Br$\gamma$ to Pa$\alpha$ line ratio. The large error of the latter value is due to the small value of the differential extinction between the two lines together with a 10\% uncertainty in the flux calibration. From the fluxes of Brackett-series lines given in Sugai et al. (1999) we obtain values of the extinction to source C of between $A_V=8\,$mag and $A_V = 12\,$mag. For source B1 the extinction does not seem to vary in such a dramatic way, but it still increases toward longer 
wavelengths, with a maximum value of $A_V = 17\,$mag. 

The differing values as a function of wavelength are the usual indication that 
the extinction is either patchy or mixed with the emitting sources along the 
line of sight (or both). In confirmation, for C1+C2 patchy extinction is seen 
from the optical/near infrared color map (see next section). In addition, our 
data indicate that the extinction increases for many sources as the aperture 
is reduced in size. In such a highly-obscured and complex environment we 
are not probing the same regions in the optical and infrared even when 
the same aperture sizes are used. This conclusion was already implicit 
in the differing morphology of the continuum images between these spectral 
regions (Section 3.1). 

The relative distribution of gas and dust appears to be similar for 
both molecular and ionized hydrogen. Our estimates from near infrared 
hydrogen recombination lines and a foreground screen are in good 
agreement with the values derived from the near  
infrared molecular hydrogen lines and a foreground screen by Kulesa 
et al. (1999) which are: $A_V = 23\,$mag for source A and $A_V = 15\,$mag 
for both sources B1 and C.

\subsubsection{Extinction to the stars}

We can obtain a qualitative idea about the differential extinction 
from the color 
maps presented in Figure~9 (infrared-infrared $H-K$ and optical-infrared 
$M_{\rm F606W}-H$). Because the WFPC2 F606W filter contains H$\alpha$, the 
optical to near-infrared color maps will show in great contrast 
bright unobscured H\,{\sc ii} regions. Both color maps confirm that  
the extinction to the nucleus of IC~694 is very high. It is interesting that 
the $H-K$ map shows a similar morphology to the molecular hydrogen emission 
(the butterfly pattern). The  $M_{\rm F606W}-H$ color map of region 
C shows patchy extinction with a very red embedded source surrounded by bright
H\,{\sc ii} regions. The ground-based  H$\alpha$ image shows very bright 
emission from region C, which suggests that the H$\alpha$ emission is 
dominated by the bright outer H\,{\sc ii} regions, whereas the near-infrared 
recombination lines trace the emission from the  very red embedded source. 
Near B1 we can clearly see the bright H\,{\sc ii} region associated with 
source B16, whereas the other sources 
(B1, and B11 through B15) seem to be obscured by a dust screen, rather 
than patchy extinction.

\begin{figure*}
\figurenum{9}
\plotfiddle{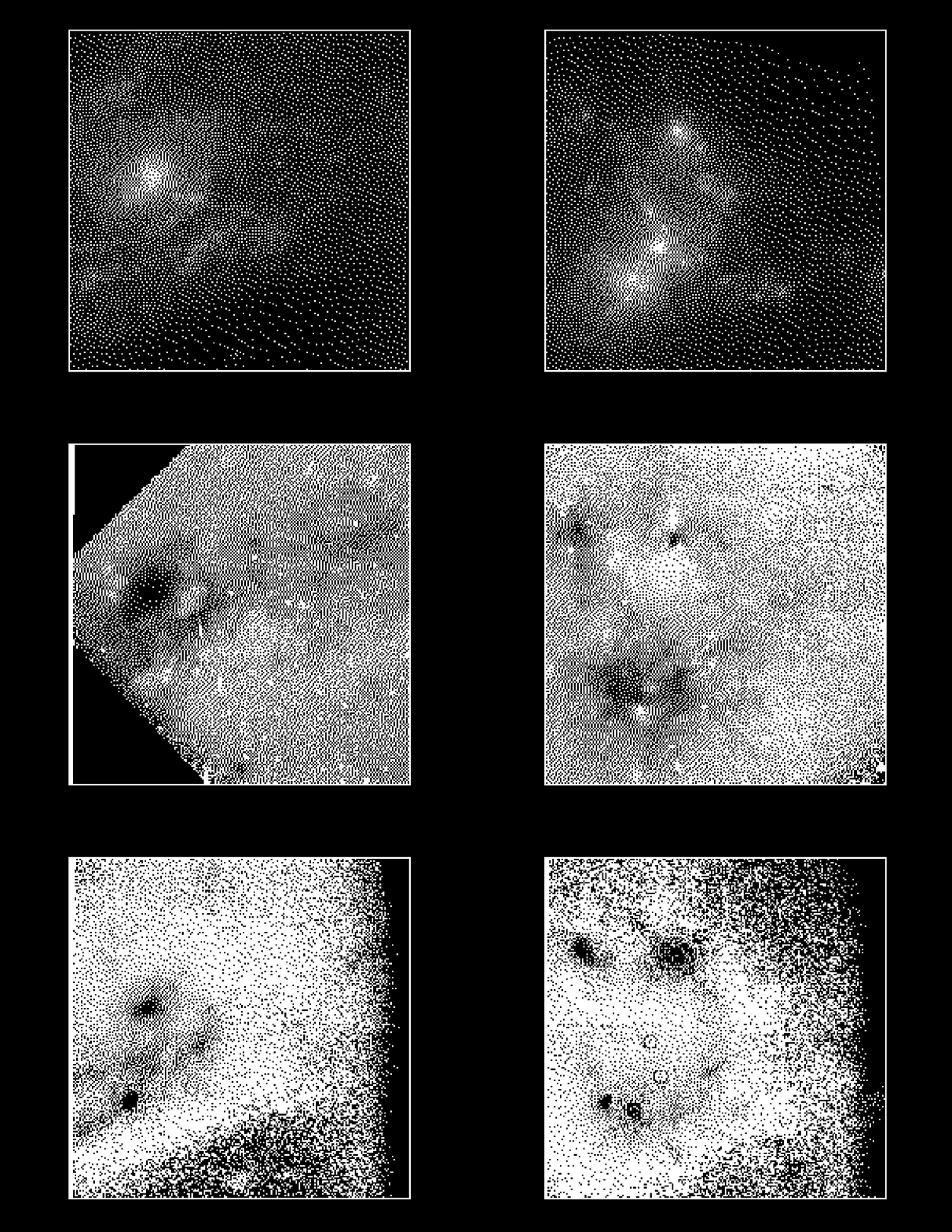}{425pt}{0}{50}{50}{-130}{50}
\vspace{0cm}
\caption{The left panels are IC~694 and the right panels 
are NGC~3690. From top to bottom, $H$-band images, 
the WFPC2 F606W / NIC2 F160W color maps (middle) and the 
NIC2 $H-K$ color maps. Orientation north up, east to the
left. The field of view of all the images is
$19.5\arcsec \times 19.5\arcsec$.}
\end{figure*}

A more quantitative estimate of the extinction to the stars can be 
obtained from the colors. The optical-infrared color $R-J$ is sensitive to age affects and the sources in Arp~299 appear to be young. We therefore use 
the $J-H$ and $H-K$ colors, which are relatively unaffected 
by age. As found for the extinction to the gas, when a simple foreground 
screen model is assumed, the values of the extinction increase from the 
$J-H$ values to the longer wavelength 
$H-K$ ones. This effect is shown in the first part of Table~7, where 
we give the extinction inferred from dereddening $J-H$  to the color of 
a normal stellar population ($J-H = 0.7$) along with the corrected 
$H-K$ and (CO)$_{\rm NICMOS}$ values using that extinction. For the 
F237M fluxes we interpolated the extinction between the values for 
$A_K$ and $A_L$  given in Rieke \& Lebofsky (1985). The dereddened 
$H-K$ colors are still very red, but the
question remains whether this is the effect of hot dust emission and/or 
differential extinction. In the second part of Table~7 we give the same 
quantities but corrected for the highest possible extinction derived from 
the hydrogen recombination lines.

We can distinguish between these extreme values of the extinction
by computing (CO)$_{\rm NICMOS}$ versus $H-K$ as a function of the 
properties of any emission by hot dust. Figure~10 shows the results 
for sources A, B1, B2 and  C1. 
The open star symbols are the values obtained for the extinction derived from 
the observed $J-H$ color, whereas the filled star symbols are the 
values obtained 
using the highest extinction derived from the hydrogen recombination lines. 
The solid lines represent mixing curves of a normal stellar population and 
hot dust with different temperatures. For the normal stellar population we 
integrate the spectrum of an elliptical galaxy redshifted to $z=0.01$ 
using the filter transmission curves and detector response functions.

The $H-K$ and (CO)$_{\rm NICMOS}$ values 
were obtained comparing to the values for the standard
star P330E as described in Section~3.3. 
For the redshifted elliptical galaxy we obtain $H-K = 0.23$ and 
$({\rm CO})_{\rm NICMOS}=0.13$. The dashed lines are the locii of the ratio  
$(f_{\rm dust}/f_{\rm total})_K$  with values 0.2, 0.4, 0.6, 0.8 and 1. 
These mixing curves are computed by adding different dust contributions 
(in the form of a blackbody) to the elliptical galaxy. Figure~10 shows that 
reproducing the (CO)$_{\rm NICMOS}$ indices and the $H-K$ colors with the 
extinction obtained by
dereddening the $J-H$ colors would require dust temperatures higher
than $T=1000\,$K for all sources. Such behavior is atypical 
of starbursts (e.g., Rieke et al. 1993).

\begin{figure*}
\figurenum{10}
\plotfiddle{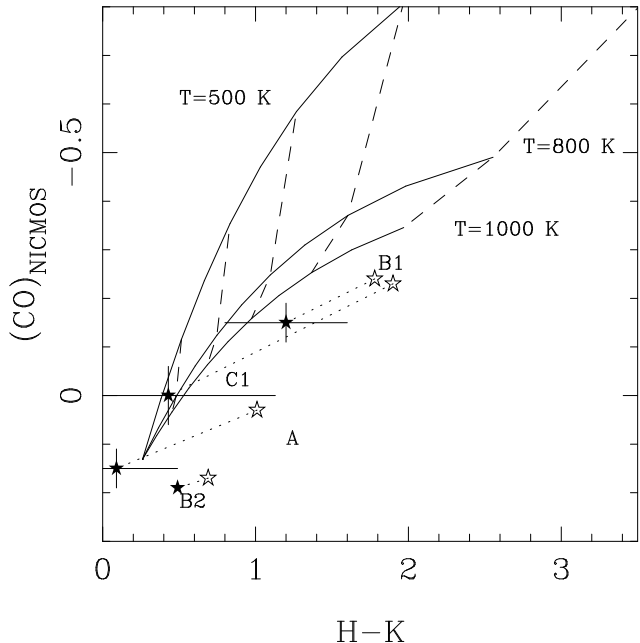}{425pt}{0}{70}{70}{-180}{160}
\vspace{-6cm}
\caption{(CO)$_{\rm NICMOS}$ versus $H-K$. The solid 
lines represent mixing curves of hot dust and a normal stellar population
redshifted to $z=0.010$
($H-K = 0.23$ and $({\rm CO})_{\rm NICMOS}=0.13$), 
whereas the  dashed lines are the
fraction of hot dust emission in the $K$-band: 
$(f_{\rm dust}/f_{\rm total})_K = 0.2,\, 0.4,\, 0.6,\, 0.8$ and 1 (from
bottom to top).
The open stars are the values for $H-K$ and $({\rm CO})_{\rm NICMOS}$
corrected for the extinction derived from the $J-H$ color, whereas 
the filled stars are the
values for $H-K$ and $({\rm CO})_{\rm NICMOS}$ corrected for the
(highest) extinction
derived from  the Br$\gamma$/Pa$\alpha$ line ratio.}
\end{figure*}

An independent approach to the same issue can be found in Satyapal et al. 
(1999). They use measurements of the 3.29$\mu$m PAH emission feature and a 
standard ratio of feature strength to hot dust continuum to conclude that 
hot dust can contribute only 7, 3, and 12\% respectively to the K-band 
fluxes of sources A, B, and C. 

The other possibility is that there is differential extinction to the
stars, as we have already found for the gas.  Using the highest values
for the gas extinctions (Table~6), we obtain more reasonable
temperatures for the dust together with a smaller dust contribution to
the $K$-band flux.  Indeed, only source B1 needs a dust contribution
of the order of 40\%, whereas the other sources require $(f_{\rm
dust}/f_{\rm total})_K < 0.2$. The fact that using the highest
extinctions derived for the gas does produce reasonable values for the
dereddened values of $H-K$ and (CO)$_{\rm NICMOS}$ suggests that the 
extinction to the gas and to the stars is similar (at least in the near-infrared). The general tendency for the extinction in the gas to be about two 
times higher than that to the stars (e.g., Calzetti, Kinney, \& 
Storchi-Bergmann 1994) does not appear to hold for this galaxy. 

A more realistic approach to the treatment of the 
extinction should include a model
in which stars and dust are mixed (e.g., Witt, Thronson, \&
Capuano 1992; Witt \& Gordon 1999). We used the recently published models
of Witt \& Gordon (1999) to evaluate the extinction to source A. We find
that the $H-K$ color excess can be only reproduced with their shell models,
using a $K$-band extinction of $A_K \simeq 1.8$. This is equivalent
to the value used to deredden the  $K$-band magnitude of component A 
when using the dust screen model (see Section~4.5), confirming 
the assumption that the system 
becomes roughly optically thin near the $K$-band. Note  
that in Witt \& Gordon (1999) models, the 
relation between $A_K$ and $A_V$ is model dependent.

Finally since $(f_{\rm dust}/f_{\rm total})_K$ 
for source C appears to be relatively small, the lack of CO bands 
is not due 
to dilution by hot dust emission, but rather due to the youth of 
the stellar 
population that dominates the $K$-band light. This conclusion is 
confirmed by 
the lack of CO bands in the $H$-band also (see spectrum in Vanzi 
et al. 1998). 

\section{STAR FORMATION PROPERTIES}
We can distinguish at least six star forming environments in the Arp~299 
system: 1.) Component A, the nucleus of IC 694; 2.) component B1; 3.) 
B2; 4.) C; 5.) C'; and 6.) additional bright H\,{\sc ii} regions 
identified individually in our Pa$\alpha$ images. In this section we 
will make use of evolutionary synthesis models (Rieke et al. 1993; 
Engelbracht et al. 1996, 1998) and other arguments to study the star 
formation properties of these regions. We will assume a truncated 
Salpeter IMF\footnote {$\phi(m){\rm d}m \propto m^{-2.35}{\rm d}m$ 
between 1 and $80\,{\rm M}_\odot$, and $\phi(m){\rm d}m \propto 
m^{-1}{\rm d}m$ between 0.1 and $1\,{\rm M}_\odot$.}. Normalized to 
the same total mass, this IMF is virtually identical to IMF8 found 
by Rieke et al. (1993) to give a good fit to the starburst properties 
of M82. We first consider the parameters that allow us to deduce the 
star forming properties of these sources.

\subsection{Luminosity}
Since Arp~299 is an infrared luminous galaxy, we can assume that most of
the bolometric luminosity is emitted in the infrared. An estimate of the fraction of the total infrared luminosity from each component can be 
obtained from ground-based mid-infrared observations  of the system 
(Gehrz et al. 1983; Wynn-Williams et al. 1991; Keto et al. 1996). The 
relative fluxes at $10\,\mu$m, $12\,\mu$m, $20\,\mu$m and $32\,\mu$m 
through a 6\arcsec-diameter \ aperture for each component are: ${\rm A} 
= 27\% \rightarrow 35\% \rightarrow 42\% \rightarrow 63\%$, ${\rm B1} = 
55\% \rightarrow 53\% \rightarrow 40\% \rightarrow 30\%$,  and 
${\rm C + C'} = 17\% \rightarrow 12\% \rightarrow 18\% \rightarrow 7\%$. 
We show below that about 10\% of the total ionizing luminosity is 
generated in supergiant H\,{\sc ii} regions that would not be 
detectable individually in the ground-based data. Hence we can assume 
that $\simeq 50\%$ of the total infrared luminosity of the system is 
concentrated at A (a similar fraction was derived by Joy et al. 1989 
for mid-infrared and IRAS far-infrared data), $\simeq 27\%$ is 
concentrated at component B1 (+ B2) of NGC~3690, $\simeq 13\%$ is at 
C+C' area, and $\simeq 10\%$ in the supergiant H\,{\sc ii} regions. From the 
mid-infrared maps of Keto et al. (1996), it can be inferred that
 most of the emission for the area B1+B2 comes actually from component
 B1. For our model fitting we will take the bolometric luminosities 
given in Table~8 as lower limits, to allow for any escaping energy.

\subsection{Mass}
An estimate of the available mass is useful as an input to the synthesis models. Shier et al. (1996) derived a mass of $5.6 \pm 1.8 \times 10^9\,{\rm M}_\odot$ for IC~694 from the velocity dispersion measured for the 2.3$\mu$m CO band. This mass is in reasonable agreement (but is larger than) the determination from radio emission lines (Zhao et al. 1997). For component B2 of NGC~3690 the dynamical mass inferred by Shier et al. (1996) is $0.6 \times 10^9\,{\rm M}_\odot$, whereas the shallow CO index of B1 did not allow them to get a reliable estimate for this component. We face a similar 
problem with source C, with its lack of CO bands.

Starburst luminosity varies more slowly than other parameters such as 
ionizing flux. Since it appears that some of the star forming regions in this system are not in nuclear potential wells, the only useful mass associated with them is the actual mass associated with the local star formation. To estimate the masses of B1 and C where there are no dynamical measurements, we assume the mass-to-light ratio derived for the starburst model of component A, $M/L_{\rm IR} = 0.004$. The masses of each component are given in Table~8. For B1 we subtracted the dynamical mass of B2. 

The mass of molecular gas in the interacting system Arp~299
has been estimated from radio CO observations assuming a standard Galactic CO to H$_2$ conversion (see for instance Sargent \& Scoville 1991). These estimates obtain the troubling result that the mass of molecular gas accounts for most of the dynamical mass ($M_{\rm gas} \simeq 0.7\times M_{\rm dyn}$). As discussed  by Shier et al. (1994) the standard conversion factor is probably a factor of 3 to 10 too high for infrared luminous galaxies. The work of Shier et al. (1994) has been questioned by Scoville, Yun, \& Bryant (1997) and Solomon et al. (1997) on the basis that $2\,\mu$m spectroscopy will not penetrate the interstellar
dust in these systems and hence will underestimate the dynamical mass. 
However, Zhao et al. (1997) used the H92$\alpha$ line to measure a dynamical 
mass for IC~694 less than, but consistent with, that obtained by Shier et 
al. (1994), supporting the general validity of the arguments of Shier et 
al. (1994). Braine \& Dumke (1998) have recently estimated H$_2$ masses from the mm-wave thermal emission of interstellar dust in Arp 299 and similar systems and find, in agreement with Shier et al., that the true H$_2$ masses are probably factors of 3 to 10 lower than given by the standard CO conversion. We therefore assume for all the nuclei that the conversion to $M({\rm H}_2)$ is 30\% of the standard value. The molecular mass in A and probably B2 is then about 15\% of the total mass, consistent with theoretical expectations for when molecular gas in a galaxy nucleus would tend to break up into star forming clouds (Wada \& Habe 1992, Bekki 1995). 

\subsection{The Number of Ionizing Photons}
Zhao et al. (1997) used observations of the radio hydrogen 
recombination line H92$\alpha$ to model the physical conditions of the 
H\,{\sc ii} regions in components A and B1 of Arp~299. Their modeling has
a number of free parameters (such as size of the H\,{\sc ii} region, electron
density, etc). The advantage in using this radio line is that it is not affected by extinction, although the estimated number of ionizing photons is slightly dependent on the model parameters (see Zhao et al. 1997 for a discussion).  The resulting uncertainties do not allow them to distinguish between 
two extreme models involving either a small number (a few tens) of very 
luminous H\,{\sc ii} regions with typical sizes of 25\,pc, or a large 
number ($\simeq 10^5$) of modest luminosity H\,{\sc ii} regions of very 
compact sizes ($\simeq 0.08\,$pc). 

The values derived by these authors are $N_{\rm Ly} = 3.8 - 4.7 \times 
10^{54}\,$s$^{-1}$ for IC~694 (component A). These values correspond to 
those from the infrared lines if $16\,{\rm mag} < A_V 
< 18\,{\rm mag}$. For component B1 of 
NGC~3690, the fits of Zhao et al. yield $N_{\rm Ly} = 1.5 - 8.8 \times 
10^{53}\,$s$^{-1}$. The larger value holds for a small number of large 
H\,{\sc ii} regions, which is as indicated by our Pa$\alpha$ imaging 
and is particularly likely for source B1 because of its compact size 
in our images. It yields an estimate of A$_V$ $\sim11\,$mag. 

Within the errors, these extinction values agree with those obtained 
from the longer wavelength infrared lines. They are significantly 
larger than would have been deduced from the Balmer decrement or even 
from the shorter wavelength infrared lines such as Pa$\beta$ 
($\lambda = 1.28\,\mu$m). We adopt values consistent with both the 
radio and infrared determinations: $A_V = 17\,$mag for A and 
$A_V = 11\,$mag 
for B1. Unfortunately Zhao et al. (1997) did not model component C. 
From our data and those of Sugai et al. (1999), we will adopt 
$A_V = 15\,$mag. In general, our arguments regarding the extinction behavior 
have close quantitative agreement with the work of Sugai et al. (1999). 
In Table~8 we give the extinction-corrected values for number of 
ionizing photons ($N_{\rm Ly}$). 

An independent test can be made from the free-free radio continuum. 
In the case of M82, Puxley et al. (1989) have compared this continuum 
with the strength of the H53$\alpha$ line, which is much less subject 
to modeling uncertainties in converting to $N_{\rm Ly}$ than is H92$\alpha$. 
We therefore have confidence that the strength of the free-free 
continuum is correctly related to N$_{Ly}$ in M82. By correcting 
for distance, we predict strengths of 14, 3, and 11 mJy for the 
free-free emission at 3mm in components A, B, and C, respectively, 
of Arp~299. The measured fluxes at this wavelength are 17 $\pm$ 2, 5 
$\pm$ 2, and 9 $\pm$ 2 mJy respectively (Aalto et al. 1997). The 
agreement is excellent. 

For component A, our data, those of Sugai et al. (1999), the 
comparison with Zhao et al. (1997), and the free-free radio flux are 
in excellent agreement but strongly disagree with the recent conclusion 
by Satyapal et al. (1999). The latter reference uses an extinction 
of only A$_V$ = 6 and deduces that the Lyman continuum flux is 
three times smaller than obtained by the other four approaches. 
Satyapal et al. acknowledge the discrepancy with the H92$\alpha$ 
flux, but they adopt the smaller ionizing fluxes suggested by their 
extinction levels for their modeling. Since our improved infrared 
data agree with the higher extinction levels, we believe their 
values are underestimated due to their use of Pa$\beta$ in the extinction 
estimate. Evidently, optical depth effects at 1.28$\mu$m are 
significant. Satyapal et al. adopted $A_V = 2\,$mag for component C, 
whereas both our data and those of Sugai et al. indicate a much 
larger value, $A_V > 10\,$mag. The resulting underestimation of the 
ionizing fluxes undermines the modeling of Satyapal et al. (1999) 
for these components. 

Although for simplicity we (and Satyapal et al.) have used a foreground 
screen model for the extinction, the agreement with Zhao et al. (1997) 
is only achieved with the 'high' extinction values from the long 
wavelength recombination lines. This behavior demonstrates that the 
dust and gas (and probably the stars) are mixed, and that the system 
becomes roughly optically thin near the $K$-band.

\subsection{The Supernova Rate}
The supernova rate in starburst galaxies can be estimated from radio
observations (assuming that the non-thermal radio flux is synchrotron
emission produced in shocks associated with supernovae), and
from  near-infrared [Fe\,{\sc ii}] emission lines  using the calibration
obtained for the starburst galaxy M82 (Vanzi \& Rieke 1997; Alonso-Herrero 
et al. 1999). Using the 20\,cm radio data of Gehrz et al. (1983) for 
the apertures which most closely match ours, and the relation between 
5 GHz flux density and supernova rate measured for M82 by Huang 
et al. (1994), we obtain 
rates of 0.42, 0.11, and $0.06\,$yr$^{-1}$ for sources A, B, and C 
respectively. The Condon \& Yin (1990) calibration for the supernova rate 
from Galactic supernovae yields similar values. However, the radio 
spectrum of source A is too flat near 20 cm to match typical 
supernovae; we believe it is affected by free-free absorption as 
previously found for M82 (Rieke et al. 1980). We use the relative 
nuclear flux densities at 8.3 GHz from Zhao et al. (1997) to 
correct for this effect. We arrive at a best estimate of the 
supernova rates of 0.65, 0.11, and 0.06 yr$^{-1}$, respectively. All 
of these values may be slightly high because free-free emission is 
expected to contribute at the $\sim$ 10-20\% level to the observed 
radio fluxes. 

We can also use the newly determined calibration between
the [Fe\,{\sc ii}]$1.644\,\mu$m luminosity and the supernova rate for M82
(Alonso-Herrero et al. 1999), which gives about a two times lower supernova 
rate than the previous best calibration (Vanzi \& Rieke 1997). 
Correcting the observed [Fe\,{\sc ii}] fluxes for the appropriate 
extinctions, we derive supernova rates of 0.52, 0.06 and 
$0.06\,{\rm yr}^{-1}$ for A, B1 and C (extinction $A=15\,$mag) 
in apertures of $2.4\arcsec \times 4.6\arcsec$, $1.2\arcsec \times 
1.2\arcsec$, and $1.2\arcsec \times 1.2\arcsec$. The agreement with 
the radio-based estimate is quite good; for the model fitting we use 
the radio results, but there would be no change if we had used [Fe\,{\sc ii}].

\begin{deluxetable}{lcccccccc}
\tablefontsize{\footnotesize}
\tablewidth{16cm}
\tablecaption{Properties of the components of Arp~299.}
\tablehead{\colhead{Source} & \colhead{Mass} & \colhead{A$_V$} &
\colhead{$L_{\rm IR}$} & \colhead{$N_{\rm Ly}$(total)} &
\colhead{$K$} & CO & SNr & EW(Pa$\alpha$) \nl
& \colhead{(M$_\odot$)} & \colhead{(mag)} &
\colhead{(s$^{-1}$)} & \colhead{(s$^{-1}$)} & \colhead{(mag)} &
& (yr$^{-1}$) & (\AA)\\
\colhead{(1)} &\colhead{(2)} & \colhead{(3)} & \colhead{(4)} &
\colhead{(5)} & \colhead{(6)} & \colhead{(7)} & \colhead{(8)} &
\colhead{(9)}}

\startdata
A & $5.8\times 10^9$ & 17 & $2.5 \times 10^{11}$ &
$4 \times 10^{54}$ & $-23.87$ &
0.17 & 0.65 & 220 \nl
B1  & $> 0.4 \times 10^9$ & 11 & $<1.4 \times 10^{11}$ &
$9 \times 10^{53}$ & $-22.64$ &
-0.15 & 0.12 & 100 \nl
B2  & $0.6 \times 10^9$ & 3.5 & -- & -- &
$-21.76$ & 0.19 & -- & -- \nl
C  &  $>0.2 \times 10^9$ & 15 & $0.5 \times 10^{11}$ &
$3 \times 10^{54}$ & $-22.71$ &
0.00 & 0.06 & 500 \nl
C' &  $>0.1 \times 10^9$ & 10 &
$0.25 \times 10^{11}$ & $3 \times 10^{53}$ & $-18.98$ &
-- & 0.03 & 1500 \nl
SG H\,{\sc ii} & $\sim 0.02 \times 10^9$ & -- &
$0.2 \times 10^{11}$ & $>2.2 \times 10^{53}$ & -- &
-- & -- & -- \nl

\enddata
\tablecomments{(1) SG H\,{\sc ii} represents the sum of the 
contributions from the 19 most luminous HII regions in the system. 
No extinction corrections have been applied for them. (2) Dynamical 
mass for A and B1 from Shier et al.
(1996). Masses for remaining components represent forming stellar 
mass to produce observed infrared luminosity. (3) Assumed foreground 
extinction level. 
(4) Estimated infrared ($8-1000\,\mu$m) luminosity for
each component. (5) Total number of ionizing photons corrected 
for extinction. (6) Extinction corrected $K$-band absolute magnitude 
for the small apertures ($2.4" \times 4.6"$ for Source A, 
$1.2" \times 1.2"$ for B1 and C, $2" \times 2"$ for C'). (7) NICMOS 
CO index. (8) Supernova rate for small apertures from 
radio. (9) Equivalent width of 
Pa$\alpha$ for the small apertures.}
\end{deluxetable}

\begin{figure*}
\figurenum{11a}
\plotfiddle{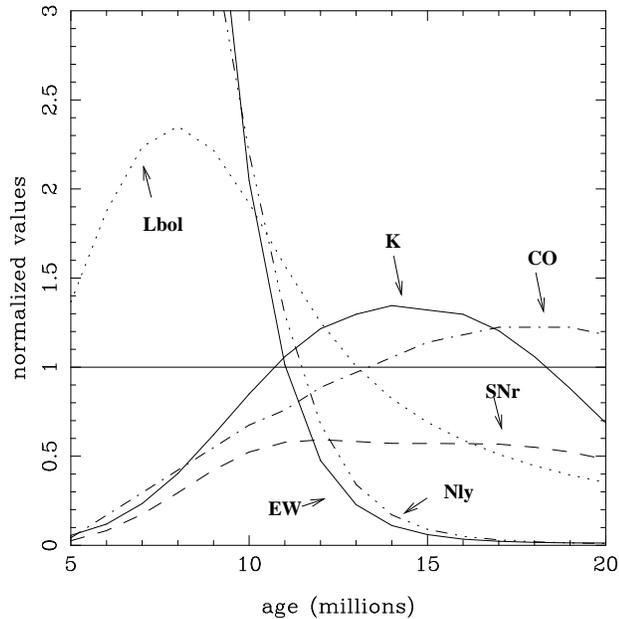}{425pt}{-90}{70}{70}{-100}{460}
\vspace{-8cm}
\caption{Evolutionary synthesis model fit for source A and 
a truncated Salpeter IMF. The extinction
to the gas is assumed to be the same as the extinction to the stars,
$A_V = 17\,$mag. The curves are as follows. Solid line (declining) the
EW of Pa$\alpha$. Solid line $K$-band luminosity. Dotted line the bolometric
luminosity. The dashed line the supernova rate. The dot-dash line the
CO index. The dot-dot-dot-dash line the number of ionizing photons.}
\end{figure*}

\subsection{Results from the evolutionary synthesis models}
We can interpret the starburst properties with evolutionary synthesis models. 
The observables that we will try to reproduce with these 
models are the bolometric luminosity $L_{\rm bol}$, the $K$-band luminosity, 
the number of ionizing photons, the CO index, the supernova rate and
the EW of Pa$\alpha$. The apertures for all the parameters are matched 
to the infrared spectroscopy. As in Rieke et al. (1993) we normalize the 
output of the evolutionary synthesis models to the observed values for each 
component; we will plot the evolution in time of the six quantities so that 
our target quantity will be unity. The mass-to-light ratios for both $K$-band 
luminosity  and the bolometric luminosity given in Engelbracht et al. (1996) 
are used to estimate the contributions of the old underlying population to the
total $K$ band and bolometric luminosities. For simplicity, we assume a single burst of star formation with a Gaussian FWHM of $5 \times 10^6\,$yr. The peak of star formation occurs at $5 \times 10^6\,$yr after the beginning of the burst.

\begin{figure*}
\figurenum{11b}
\plotfiddle{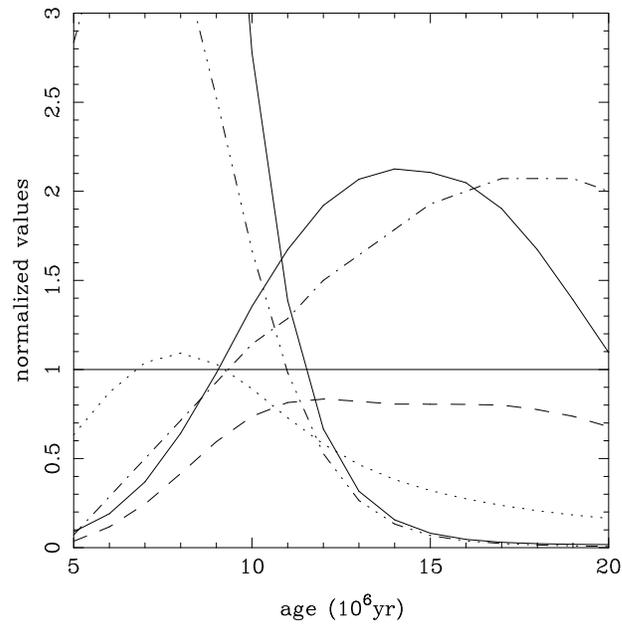}{425pt}{0}{70}{70}{-180}{220}
\vspace{-8.5cm}
\caption{Evolutionary synthesis model fit for 
source B1 and a truncated Salpeter IMF.
The extinction  is assumed to be $A_V  = 11\,$mag.}
\end{figure*}

\begin{figure*}
\figurenum{11c}
\plotfiddle{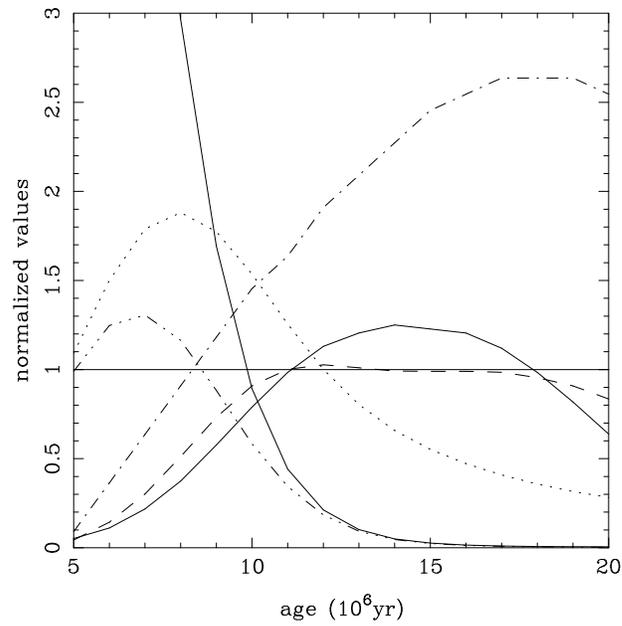}{425pt}{0}{70}{70}{-180}{280}
\vspace{-10cm}
\caption{Evolutionary synthesis model fit for source C and a
truncated Salpeter IMF. The extinction
to the gas is assumed to be the same as the extinction to the stars,
$A_V = 15\,$mag.}
\end{figure*}

\begin{deluxetable}{lcccc}
\tablefontsize{\footnotesize}
\tablewidth{12cm}
\tablecaption{Synthesis Fit to C'.}
\tablehead{\colhead{} & \colhead{$L_{\rm IR}$} &
\colhead{$N_{\rm Ly}$(total)} &
\colhead{$K$} & SNr \nl
& \colhead{(L$_\odot$)} & \colhead{(s$^{-1}$)} & \colhead{(mag)} 
& (yr$^{-1}$) \\
\colhead{(1)} &\colhead{(2)} & \colhead{(3)} & \colhead{(4)} &
\colhead{(5)}}
\startdata
Observations &  $0.25 \times 10^{11}$ &
$3 \times 10^{53}$ & $-18.98$ &
0.03 \nl
Model &  $0.3 \times 10^{11}$ &
$4 \times 10^{53}$ & $-19.0 $ &
0.025 \nl
\enddata
\end{deluxetable}

{\it Source A}.\\
We assume an extinction of  $A_V = 17\,$mag. The $K$ band luminosity
and the $N_{\rm Ly}$ are corrected for extinction using a foreground dust 
screen. Because we find that the source becomes roughly optically thin
at $K$, this procedure should give reasonably accurate results. For the 
CO index, we take the value from Shier et al. (1994), which is measured 
spectroscopically and can be converted accurately to the photometric 
index used in the models. The fit for the observables of source A 
(Figure~11.a) uses a total mass in newly formed stars of 
$7 \times 10^8\,$M$_\odot$ ($\simeq 12\%$ of the dynamical mass) 
and yields  an age for the star formation episode of $6-8\times 
10^6\,$yr (after the peak of star formation). There is a modest 
discrepancy in the supernova rate.
Measured from the rising half power of the assumed 
Gaussian in the star formation rate, the age of the burst is then 
$\sim$ 11 Myr.

{\it Source B1}\\
For the assumed level of the extinction ($A_V =11\,$mag), the 
contribution of hot dust emission to the observed $K$ luminosity is 
$(f_{\rm dust}/f_{\rm total})_K \simeq 0.5$,
which needs to be subtracted to get the stellar $K$-band luminosity. 
The observed CO index and equivalent width of Pa$\alpha$ also need to be
corrected for the dust emission contribution to the continuum. We 
assume that approximately 70\% of the bolometric luminosity of B1+B2 
corresponds to B1. The best fit for source B1 is presented in 
Figure~11.b. The observed values are well
fitted at an age between 4.5 and $7\,$Myr after the peak of star formation. 
The weak CO absorption suggests an age near the lower end of this 
range, $\sim$ 5 Myr after the peak, or $\sim$ 7.5 Myr from the 
rising half power of the assumed Gaussian in the star formation rate.

{\it Source C}\\
Sugai et al. (1999) detect the $1.70\,\mu$m He\,{\sc i} line at 
about 30\% the 
strength of Br 10. Thus, the ionizing flux from this region is dominated 
by stars with temperatures above 40,000\,K (Vanzi et al. 1996). We have 
already noted the lack of first overtone CO bands from red giant and 
supergiant stars. Together, these observations require that the starburst 
in this component be extremely young, dominated by hot stars which have 
not yet evolved significantly into the red supergiant phase. However, 
the presence of significant supernova activity, as shown both by the 
[Fe\,{\sc ii}] emission and nonthermal radio flux, places a strong lower 
limit of $\sim$ 3.5 Myr on the age of the starburst.

As a first case for synthesis modeling, we consider a value for the 
extinction to the gas and to the stars of $A_V = 15\,$mag. From 
Figure~10 we get that $(f_{\rm dust}/f_{\rm total})_K \simeq 0.5$.  
Again the observed
CO index (value taken from Ridgway et al. 1994 and the conversion
from Doyon et al. 1994 as above) and the equivalent width of 
Pa$\alpha$ are corrected for the dust contribution to the continuum.  
The best fit for the truncated Salpeter IMF is presented in Figure~11.c. 
It corresponds to between
3 and 6\,Myr after the peak of star formation. Only the younger of 
these ages appears possible given the evidence of extreme youth 
from the He\,{\sc i} line. The age after the rising half power point 
in star formation is then $\sim$ 6Myr.

Alternately, we can correct the observables for the highest possible 
extinction obtained from the near-infrared spectroscopy $A_V = 37\,$mag
(the dust emission contribution to the $K$-band is 
less than 20\% and the SNr given in Table~8 needs to be increased). 
We have to assume that 85\% of the total bolometric luminosity is 
unaccounted for in the observed total infrared luminosity, which argues 
against this high value of extinction. The age derived is even 
younger than for the previous fit.

{\it Source C'}\\
Component C' presents a dilemma. Its radio spectrum is nonthermal 
(Baan \& Haschick 1990) and its flux density is about 50\% of that 
of component C. However, it is very faint in the near infrared. Our 
starburst models (Figure 11) assume a Gaussian time dependence with 
a FWHM of 5\,Myr and the supernova rate leads the $2\,\mu$m flux only 
slightly. However, as the model ages, the $2\,\mu$m flux will persist, 
so the only time to fit the properties of component C' is during the 
initial period. To achieve a large ratio of radio to near infrared 
flux requires that the star formation episode be much shorter in 
duration than 5\,Myr. Rather than reconfigure our starburst program 
for an instantaneous burst, we use Leitherer \& Heckman (1995). Their 
calculations show that, in such a burst, the supernovae reach their 
typical level abruptly starting at 3.5\,Myr after the burst, while 
the red supergiants only begin to appear at about 4.5\,Myr after the 
burst. Component C' appears to be adequately explained, then, by a 
burst of nominally $\sim 3 \times 10^7\,{\rm M}_\odot$ with an age of 
4\,Myr. The excellent fit achieved is summarized in Table~9.

It is often found that nuclear starbursts can only be fitted 
assuming vigorous star formation over 5-10 Myr, as we indicated 
in this work for component A and as studied in detail for M82 by 
Rieke et al. (1993) and Satyapal et al. (1997). It is interesting 
to discover for component C', a region of similar luminosity but 
not in a potential well, that a very short duration event is 
required. A similar situation may hold for component C and possibly 
for B1 (if it is not the second galaxy nucleus), but they are not 
at the exact stage in their evolution that allows us to draw such 
a conclusion.

\begin{figure*}
\figurenum{12}
\plotfiddle{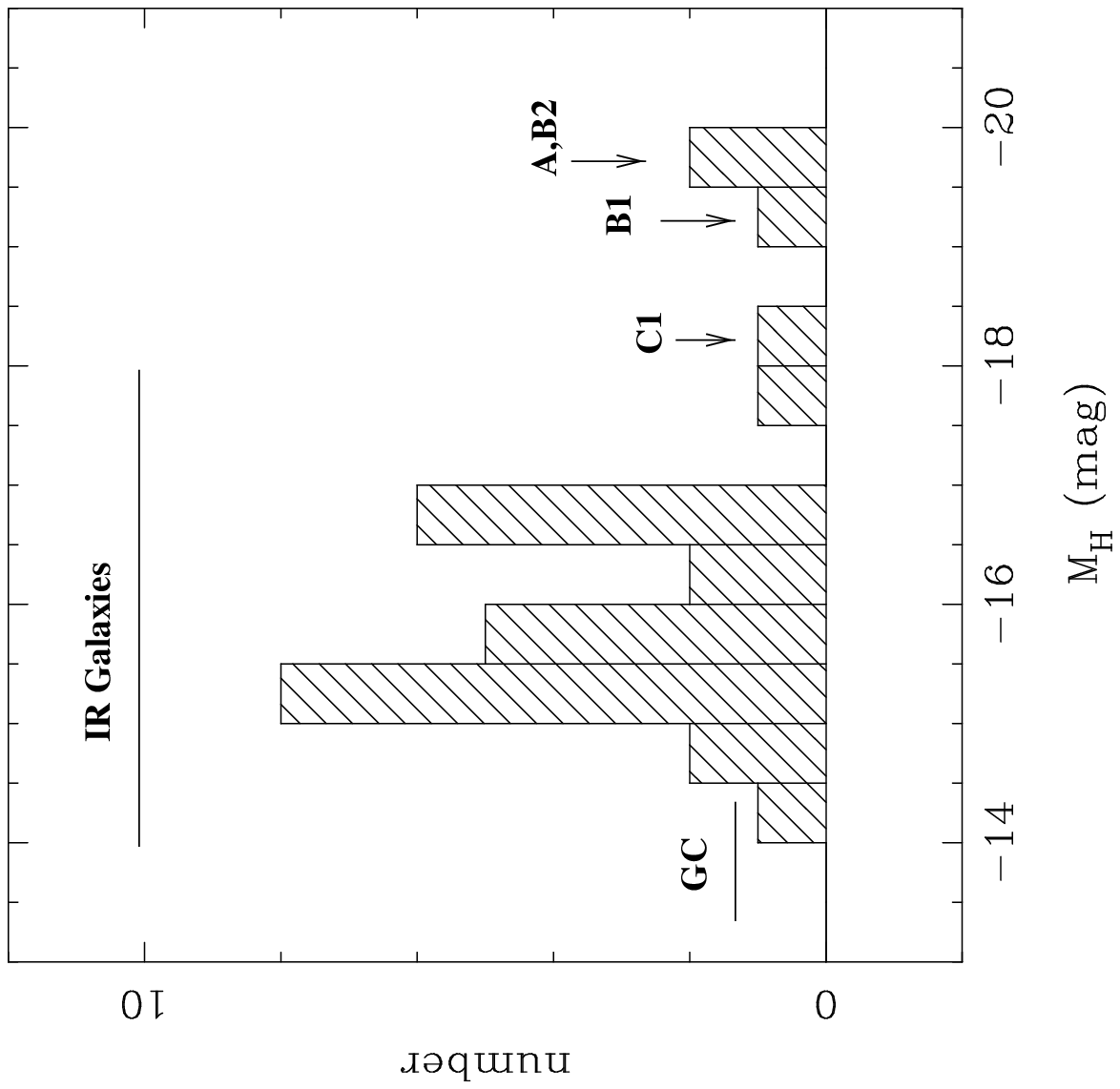}{425pt}{-90}{60}{60}{-100}{430}
\vspace{-7.5cm}
\caption{Histogram of the absolute $H$-band magnitudes of the faint
sources detected in Arp~299 (not corrected for extinction). Also shown
the  range of $H$-band magnitudes for globular
clusters and other IR luminous galaxies.}
\end{figure*}

\subsection{Other Star Forming Components}

{\it Source B2}\\
In addition to these four sources, we deduce that B2 is the site 
of recent star formation because of: 1.) its strong CO absorption 
(this work); 2.) its mid infrared excess (Gehrz et al. 1983; Keto 
et al. 1997); and 3.) its nonthermal radio emission (Gehrz et al. 
1983, Zhao et al. 1997). We can put a lower limit of $\sim$ 11 Myr 
on the age of this incident because of the very strong CO absorption 
feature, similar in depth to that of source A. As can be seen from 
Figure 11, the supernova rate for a given stellar population remains 
nearly constant from an age of $\sim$ 6 Myr to $>$ 15 Myr. Scaling
 by the radio emission, we conclude that about $3 \times 10^7 M_\odot$ 
must have been converted into stars in B2. 

{\it Supergiant H\,{\sc ii} Regions}\\
Our Pa$\alpha$ images reveal a large number of extremely luminous 
H\,{\sc ii} 
regions outside the components listed above. We have no way to correct
the Pa$\alpha$ luminosities for extinction, so the derived values are 
lower limits. We have set a lower limit of Log (Pa$\alpha$ luminosity) 
$\ge$ 38.7\,erg s$^{-1}$, implying Log (H$\alpha$ luminosity) 
$\ge$ 39.7\,erg s$^{-1}$, equivalent 
to the ionizing luminosity of 30 Doradus. We have located 19 H\,{\sc ii}
 regions 
in the Arp~299 system above this limit. Although we cannot model these 
regions in detail, we can use the evolutionary models to estimate 
roughly that each of them involves $\sim$ $10^6$ M$_\odot$ of recently 
formed stars, if the stars form according to the truncated Salpeter 
IMF and the regions are young.

We can compare this result with the study of 30 nearby galaxies in the
 H$\alpha$ line by Kennicutt, Edgar, \& Hodge (1989). Our method of 
extracting ionizing line fluxes should give values systematically 
similar or perhaps lower than used in that work (Kennicutt, private 
communication). They show that the presence of even one or two H\,{\sc ii} 
regions at the 30 Dor level is uncommon in normal galaxies. Although 
they find that the incidence of supergiant H\,{\sc ii} regions increases for 
distorted and interacting galaxies, the extreme number of such objects 
in Arp~299 is remarkable.  

Zhao et al. (1997) called attention to a region with bright 
H92$\alpha$ extending to the SE of component A. To a significant 
extent, this feature appears to be resolved by the NICMOS images 
into a few supergiant regions. However, the orientation of the 
feature in the radio maps indicates there may be additional emission 
in less spectacular condensations.

{\it Massive Stellar Clusters}\\
Much of the past star formation in Arp 299 appears to have been 
organized into luminous clusters, similar to those found in other 
interacting and highly luminous infrared galaxies. The absolute 
$H$-band magnitudes for the clusters 
in Arp~299 (not corrected for extinction) range up to $M_H=-17\,$mag
(see histogram in Figure~12). This distribution does not
include most of the low-luminosity sources, since those were not detected
in the NIC1 F110M image. We can compare with the young stellar clusters 
found in the Antennae, $M_H = -11\,$mag to $M_H = -16\,$mag 
(Whitmore \& Schweizer 1995) assuming $V-H\simeq 2$.  
The derived sizes for the stellar clusters in the Antennae are 
12\,pc (for $H_0 = 75\,$km s$^{-1}$ Mpc$^{-1}$, Whitmore \& 
Schweizer 1995), whereas our diffraction limit at $1.1\,\mu$m is 
approximately 18\,pc. In the merger galaxy
NGC~7252 young clusters range from $M_H = -16\,$mag to $M_H = -18\,$mag 
(Whitmore et al. 1993). The luminosities of 
the stellar clusters in other infrared luminous and ultraluminous galaxies 
(Scoville et al. 1999) are also similar to those in Arp 299. However, it
should be noted that the ULIRGs in Scoville et al. (1999) paper are 
more distant than Arp299, and despite the similarities in 
$H$-band luminosities, we may be probing different
spatial resolutions.

The cluster luminosities in these extreme galaxies may exceed 
the limits found in more normal conditions. For example, the 
intermediate-aged clusters in M100 
(Ryder \& Knapen 1999) have $M_H = - 12\,$mag to $M_H = -15\,$mag 
assuming ($H-K \simeq 0.2$). Since the infrared luminosities change 
only slowly (see Figure 11), we can compare directly to see that 
these values may be about 2 magnitudes lower than for ultraluminous 
galaxies. Further data are needed to confirm this trend.  

A crude estimate of the masses of the stellar clusters in Arp~299 can
be obtained using the evolutionary synthesis models presented in the
preceding
section. Although at ages $>$ 25 Myr, the luminosity will fade, it 
seems unlikely that the clusters in Arp 299 are significantly older 
than this value, given the young ages derived for the rest of the star 
formation. In Figure~13 ({\it left panel}) we present a color-magnitude 
diagram
(absolute $H$-band magnitude versus the $(H-K)$ color) for the stellar 
clusters, 
not corrected for extinction. The lines are the output of models with a 
truncated Salpeter IMF and masses of $1.5 \times 10^5\,$M$_\odot$, 
$7.5 \times 10^5\,$M$_\odot$, and $5 \times 10^6\,$M$_\odot$. The peak 
of the $M_H$ magnitude on each curve corresponds to an age of $10^7\,$yr, 
the age at which the red supergiants make the largest contribution. The 
evolution is shown from $1\times 10^6\,$yr up to $5\times 10^7\,$yr. 
The star-symbols are values for the stellar clusters. The errors are 
0.15\,mag in the colors, and $0.1\,$mag in the absolute magnitude, 
and represent the uncertainties associated with the
subtraction of the underlying galaxy (Section~3.2). Most of the stellar
clusters lie to the right of the model predictions indicating that some
extinction is present.

From Table~5 it is clear that most of the stellar clusters do not show
extremely red colors as it is the case of the bright nuclei A, B1, B2 and C1.
However the observed colors are still redder than the values predicted by
the model. The predictions of the model show variations in the $J-H$ color
of less than 0.1\,mag around $J-H =0.7$ for ages $> 10^7\,$ yr.
Since the ages of the clusters are expected to be greater
than $10^7\,$yr, one can use the observed $J-H$ color to obtain an
estimate of the foreground extinction. We find that the
extinctions are typically $A_V=1.7-4\,$mag, much lower than the extinction
to the bright sources. Figure~13 ({\it right panel}) shows how the
values of $M_H$ and $(H-K)$ corrected for extinction lie closer to the
model predictions, although they are still somewhat 
redder. The masses of these clusters are in the range of
$1.5\times 10^5-5\times 10^6\,{\rm M}_\odot$. For comparison the masses of
the stellar clusters found in other infrared luminous galaxies are
$10^5-10^7\,{\rm M}_\odot$ (Scoville et al. 1999), whereas the masses
of Galactic globular clusters are of the order of $10^5\,{\rm M}_\odot$
(van der Bergh 1995).

\begin{figure*}
\figurenum{13}
\plotfiddle{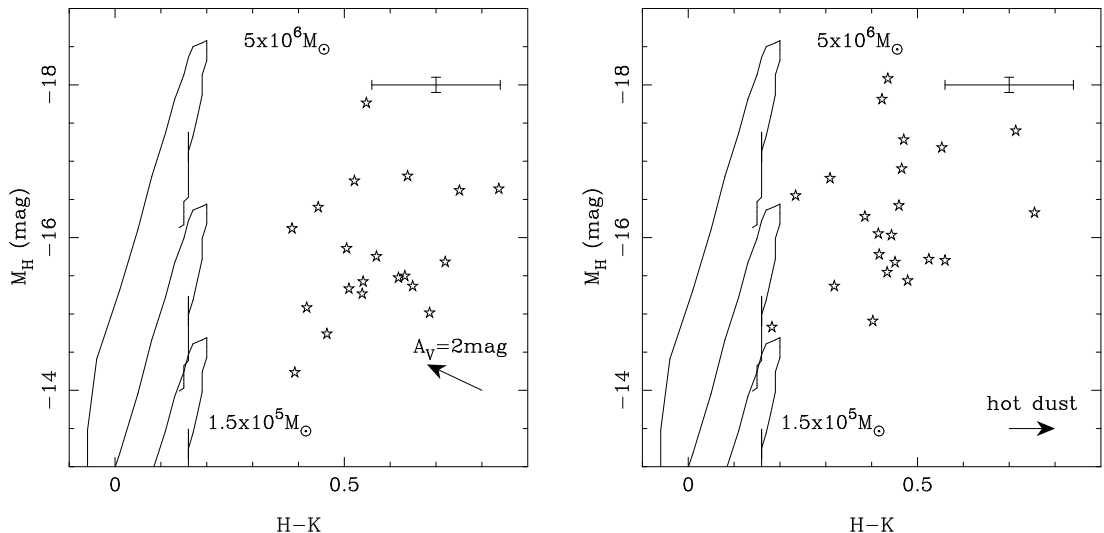}{425pt}{-90}{60}{60}{-240}{540}
\vspace{-6cm}
\caption{{\it Left panel} Absolute $H$-band magnitude versus
$(H-K)$ for the stellar clusters, not corrected for extinction. The
lines are the output of models using a truncated Salpeter IMF, 
and cluster masses of
$1.5 \times 10^5\,{\rm M}_\odot$, $7.5 \times 10^5\,{\rm M}_\odot$, and
$5 \times 10^6\,{\rm M}_\odot$. {\it Right panel} Same but using
colors and magnitudes corrected for extinction. The error bars are 
indicated in the upper right corners of the plots.}
\end{figure*}

Given the modest extinctions to the clusters, their deep CO bands 
(see Table 5) require little correction. Assuming roughly solar 
metallicity, these band strengths can only be produced by stellar 
populations of age $\sim$ 10-30 Myr. 

\subsection{An Extended Period of Elevated Star Formation}

The major starforming components in Arp 299 show a large range of ages. 
Even though some uncertainties remain regarding the ages and burst 
strengths, we find an age gradient from B2 ($>$ 11 Myr) to A ($\sim$ 
11 Myr) to B1 ($\sim$ 7.5 Myr), C ($\sim$ 5 Myr), and C' (4 Myr). 
Compared with Satyapal et al. (1999), we derive an older age for 
source A, assuming that its starburst has had a relatively long 
duration (they assumed an instantaneous burst), a similar age for 
source C, and we resolve source B into two bursts of differing ages, 
one older than and the other younger than their value. The effective star 
formation rate (mass of the newly formed stars over the FWHM of the burst) is
high for component A ($140\,{\rm M}_\odot$ yr$^{-1}$). It is moderately 
high also for B1 ($40\,\,{\rm M}_\odot$ yr$^{-1}$) and for 
C ($22\,\,{\rm M}_\odot$ yr$^{-1}$) (or $120\,\,{\rm M}_\odot$ yr$^{-1}$  
if the extinction to C is as high as 37 magnitudes). The masses in 
newly formed stars are about $7\times10^8\,{\rm M}_\odot$ for component A, 
about $1\times 10^8\,{\rm M}_\odot$ for B1,  $1\times 10^8\, 
{\rm M}_\odot$ for C, 1 $\times$ 10$^8$ ${\rm M}_\odot$ for C', and 
0.3 $\times$ 10$^8$ ${\rm M}_\odot$ for B2.

We find a similar range for star formation in very massive stellar 
clusters. We have identified 19 supergiant H\,{\sc ii} regions, and have 
deduced total stellar masses for them that are similar to the masses 
of 21 stellar clusters with -15 $>$ M$_H$ $>$ -17. The H\,{\sc ii} regions 
must be young, 5 Myr or less, whereas the strong CO absorption in 
the clusters implies typical ages $>$ 10 Myr. 

Star formation has been occurring contemporaneously throughout the 
two galaxies over the past 10-15 Myr, both within and outside of 
their nuclei. There appear to be four potentially differing 
environments for the star formation. Outside of the nuclei and along 
the spiral arms, there are massive stellar clusters and 
H\,{\sc ii} regions. 
Within the nuclear potential wells, young stars have formed in numbers 
one to two orders of magnitude greater than in these clusters. 
Although we cannot be sure of the spatial distribution of star 
formation in the nuclei now, it is plausible that these populations 
will merge into a single super-cluster, given the high density of 
young stars and the strong tidal forces that would cause clusters 
to disintegrate rapidly. In the interaction region, components C 
and C' appear to be locations where star formation has occurred on 
the scale typical of a galaxy nucleus, not an extranuclear cluster. 
Presumably, there is no nuclear potential well in this region, but 
the concentration of molecular gas in the interaction has resulted 
in an exceptionally vigorous star formation episode. This is also
found in other interacting galaxies (e.g., VV114 Frayer et al.
1999; the Antennae Vigroux et al. 1996). Finally, there 
may be examples of extreme star formation near the galaxy nuclei 
where infalling gas clouds became unstable and formed stars before 
falling all the way into the nuclear potential. One possible example 
is the region SE of the nucleus of IC 694 (component A) in the 
H92$\alpha$ measurements of Zhao et al. (1997) and partially resolved 
into supergiant H\,{\sc ii} regions in our data. Another possible example is 
components B1 and B2. It is not obvious which should be associated 
with a galaxy nucleus. However, we have found that both involve a 
larger mass of young stars than can be attributed to a supergiant 
H\,{\sc ii}
region or massive stellar cluster, so it is plausible to assume one 
is a galaxy nucleus and the other a very nearby star forming region. 

\section{FATE OF THE STARBURST REGIONS}

{\it Sources B1 \& B2}\\
The measure of stellar dynamics in B2 (Shier et al. 1996) suggests a 
mass of $6 \times 10^8$ M$_\odot$, whereas from the radio luminosity 
we have found that only about $3 \times 10^7$ M$_\odot$ is expected 
from recent star formation. Thus, this object appears to satisfy 
expectations for a modest mass galaxy nucleus. On the other hand, the 
$^{12}$CO emission is centered on B1 and has a FWHM of 260 km s$^{-1}$ 
(Aalto et al. 1997), possibly indicating a larger mass there, $\sim 3 
\times 10^9$ M$_\odot$. It is possible that the CO is subject to 
non-gravitational forces at B1, or that the stellar population at 
B2 is not gravitationally bound, so the situation of these two 
nearby concentrations is not firmly determined by the observations. 

If B1 has a mass of $3 \times 10^9$ M$_\odot$, then the Roche radius 
for B2 is $\sim$ 200 pc, about one third of the projected separation 
of the sources. In this case, B2 will be undergoing significant 
tidal evaporation and is in the process of merging into the general 
stellar distribution around B1. On the other hand, the dynamical 
mass of B2 is similar to the mass of B1 indicated for the recent 
star formation. We may be seeing two galaxy nuclei of comparable 
masses, one older and the other newly formed, orbiting each other 
at a projected distance 5-6 times the Roche radii. The dynamical 
evolution will be more complex in this situation and should develop 
more slowly, but will probably have a similar end point. 

{\it Sources C \& C'}\\
Sources C and C' are well isolated from the galaxy nuclei in the 
system - they are about 3 kpc from component A and 1.5 kpc from B, 
projected onto the sky. They do not appear to lie within any 
significant potential wells of their own, based on the lack of 
evidence for an underlying concentration of old, red stars. This 
case is particularly strong for C', given how faint it is in the 
near infrared. Both objects have low velocity dispersions in 
$^{12}$CO of 60 - 80 km s$^{-1}$ (Aalto et al. 1997). Given our 
estimates of the total mass being converted into stars (Table 8), 
the low velocity dispersion is consistent with the formation of a 
bound stellar system. The relaxation time for such a system can 
be estimated from the expression in Spitzer (1987) to be long, of 
order 10$^{12}$ yr. Thus, it would appear that C and C' represent 
the formation of stellar systems on the scale of dwarf galaxies. 
A further confirmation of this possibility comes from the star 
formation efficiency. The gas masses of both C and C' are 
$\sim$ 3 $\times$ 10$^8$ M$_\odot$ (from the $^{12}$CO measurements 
of Aalto et al. (1997) and our adjusted conversion to H$_2$ mass, 
see Section 4.2). The mass of newly formed stars is likely to be 
$ 1 \times$ 10$^8$ M$_\odot$ for both, as indicated in Table~8. 
Thus, a significant fraction of the molecular material in these 
regions has probably been turned into stars, a situation that 
favors formation of bound systems. 

The effects of tidal forces on this system can be estimated roughly from the Roche limit, which is about 500 pc relative to the nucleus of IC 694. Therefore, the stellar systems may persist for a significant lifetime, so long as they do not approach more closely to the massive nucleus of IC 694 than at present. They are perhaps best described, given their masses and potential stability, as new dwarf galaxies. 

{\it Supergiant H\,{\sc ii} Regions \& Stellar Clusters} \\
Although the giant stellar clusters seen in other starbursts have 
been described as young globular clusters, the distribution of these 
objects along spiral arms argues against a literal interpretation 
of this view. In IC 694, many of these objects lie a few Roche radii 
from the galaxy nucleus in projected distance. If the larger mass 
estimate holds for B1 in NGC 3690, a similar situation holds there. 
It is therefore plausible that the clusters that lie relatively 
close to the massive galaxy nuclei will be subject to strong tidal 
evaporation and will fade in prominence as the massive stars die 
and the less massive ones are lost into the surrounding galaxies. 
However, clusters that lie relatively far from nuclear mass 
concentrations may be reasonably stable. 

\section{CONCLUSIONS}

We have used {\it HST}/NICMOS broad-band and narrow-band images of 
Arp~299, along with archival HST data and ground-based spectroscopy 
and imaging to study this complex system. We find:

\noindent 1.) Arp 299 is dominated by widespread and powerful 
star formation. In common with other recent studies, we find no 
evidence for an AGN anywhere in the system.

\noindent 2.) As found for other infrared starburst and 
luminous and ultraluminous 
galaxies, the extinction toward the dominant infrared sources is 
extremely strong and complex. Studies using optical or near infrared 
measures of extinction (e.g., Satyapal et al. 1999) will significantly 
underestimate its importance and will deduce erroneous values 
for the critical parameters characterizing the star forming activity. 
Studies based only on UV light will face even more severe difficulties.

\noindent 3.) The infrared [Fe\,{\sc ii}] line emission predicts a rate of 
supernova activity in close agreement with that from the nonthermal 
radio fluxes, and both are in satisfactory agreement with the 
predictions of evolutionary synthesis starburst models.

The detailed morphology of the components of Arp~299 has been revealed 
with the {\it HST}/NICMOS broad-band images. The system is a 
veritable museum of interaction-induced star formation. It includes:

\noindent 1.) A massive galaxy nucleus (IC 694, source A) undergoing 
an extended-duration starburst and accounting for 
$3 \times 10^{11}\,{\rm L}_\odot$, about 50\% of the total luminosity 
of the system. 

\noindent 2.) A strongly star forming region near this nucleus, 
H92$\alpha$ SE, probably indicating a region where infalling 
molecular clouds broke up into star forming regions before 
getting all the way into the nuclear potential well.

\noindent 3.) A pair of sources (NGC 3690, B1 and B2) probably 
representing a less massive galaxy nucleus and a vigorous star 
formation episode where molecular clouds became unstable in 
its vicinity, together representing about 
$1.5 \times 10^{11}\,{\rm L}_\odot$, about 27\% of the total. 

\noindent 4.) Very young and vigorous star forming regions 
(C and C') near the interface between the two galaxies. They 
are likley to be forming gravitationally bound stellar 
systems that will eventually have the appearance of dwarf galaxies. 

\noindent 5.) A numerous population of supergiant H\,{\sc ii} regions 
and stellar clusters both near the more powerful star forming 
regions and along spiral arms. They appear to represent young 
and old phases of extreme star formation outside the galaxy nuclei. 

We have used evolutionary synthesis modeling and other arguments 
to assign ages to these various manifestations of star formation, 
finding that they have all persisted for at least $10 - 15\,$Myr. Thus, 
Arp~299 has been in a condition similar to what we observe now for 
about this period of time, with all or nearly all the varieties of 
star forming environments active. 

Arp 299 should be similar to other luminous infrared galaxies. 
The greater complexity revealed for it is primarily a result of 
its relative proximity (42\,Mpc) and our use of very high resolution 
near infrared images from NICMOS, giving a physical resolution 
of about 20\,pc.

\section*{Acknowledgments}

During the course of this work AA-H was supported by the National
Aeronautics and Space Administration on grant NAG 5-3042 through the
University of Arizona. The work was also partially supported by the
National Science Foundation under grant AST-95-29190.
We are grateful to Kevin Luhman for obtaining the optical spectra
presented in this paper, and to Craig Kulesa for providing us with data
prior to publication. We thank Karl Gordon andn Chad Engelbracht for 
helpful discussions.

{}

\end{document}